\documentclass[prd,amsmath,amssymb,superscriptaddress,floatfix,nofootinbib,10pt]{revtex4}
\usepackage{times}
\usepackage{amssymb,amsbsy,amsmath,amsfonts}
\usepackage{graphicx}
\usepackage{float}
\usepackage{color}
\usepackage{orcidlink}
\usepackage{morefloats}
\usepackage{rotating}
\usepackage{srcltx}
\usepackage{slashed}
\usepackage{multirow}
\usepackage{verbatim}
\usepackage{hyperref}
\usepackage{tabularx}
\usepackage{bm}
\allowdisplaybreaks[4]

\usepackage{bbding}
\usepackage{threeparttable}

\usepackage{mathrsfs}

\DeclareUnicodeCharacter{2212}{\ensuremath{-}}

\newcommand{\PreserveBackslash}[1]{\let\temp=\\#1\let\\=\temp}
\newcolumntype{C}[1]{>{\PreserveBackslash\centering}p{#1}}
\newcolumntype{R}[1]{>{\PreserveBackslash\raggedleft}p{#1}}
\newcolumntype{L}[1]{>{\PreserveBackslash\raggedright}p{#1}}

\begin{document}

\title{Molecular states with bottom mesons and multistrange baryons systems}

\author{Jing Song\,\orcidlink{0000-0003-3789-7504}}
\affiliation{School of Physics, Beihang University, Beijing, 102206, China}%
\affiliation{Departamento de Física Teórica and IFIC, Centro Mixto Universidad de Valencia-CSIC Institutos de Investigación de Paterna, 46071 Valencia, Spain}

\author{Yi-Yao Li}
\affiliation{
State Key Laboratory of Nuclear Physics and 
Technology, Institute of Quantum Matter, South China Normal 
University, Guangzhou 510006, China}
\affiliation{Key Laboratory of Atomic and Subatomic Structure and Quantum Control (MOE), Guangdong-Hong Kong Joint Laboratory of Quantum Matter, Guangzhou 510006, China }
\affiliation{ Guangdong Basic Research Center of Excellence for Structure and Fundamental Interactions of Matter, Guangdong Provincial Key Laboratory of Nuclear Science, Guangzhou 510006, China  }
\affiliation{Departamento de Física Teórica and IFIC, Centro Mixto Universidad de Valencia-CSIC Institutos de Investigación de Paterna, 46071 Valencia, Spain}

\author{ Eulogio Oset\,\orcidlink{ https://orcid.org/0000-0002-4462-7919}}
\email[]{oset@ific.uv.es}
\affiliation{Departamento de Física Teórica and IFIC, Centro Mixto Universidad de Valencia-CSIC Institutos de Investigación de Paterna, 46071 Valencia, Spain}
\affiliation{Department of Physics, Guangxi Normal University, Guilin 541004, China}

\begin{abstract}
We investigate molecular states formed by bottom mesons and multistrange baryons from both octet and decuplet flavor representations, using the local hidden gauge approach combined with coupled-channel Bethe-Salpeter equations. Focusing on strangeness sectors \(S = -1\) to \(-4\), we predict several bound states  in the \(S=-1, I=1/2,~3/2\) and \(S=-2, I=0\) sectors. No bound states are found in other  isospin channels, where the interaction is repulsive, nor in higher strangeness sectors. The binding energies are analyzed under different values of the cutoff regularization  parameters, providing estimates of theoretical uncertainties. This study  provides concrete predictions to support future experimental investigations  and improve understanding of heavy-flavor multistrange exotic hadrons.
\end{abstract}


\maketitle

\section{Introduction}

In recent years, advances in high-energy hadron facilities such as Belle, BaBar, BESIII, LHCb, and CMS have enabled the observation of numerous hadronic states with heavy quarks. Many of these states exhibit properties that cannot be accommodated by the traditional quark model of mesons as $q\bar{q}$ and baryons as $qqq$.
The study of exotic hadrons beyond the conventional $qqq$ and $q\bar{q}$ configurations has significantly advanced over the past two decades. Experiments have discovered a series of remarkable states including $X$, $Y$, and $Z$ mesons~\cite{Belle:2003nnu,BaBar:2005hhc,BESIII:2013ris,Belle:2013yex,Belle:2013shl,LHCb:2014zfx,LHCb:2020bwg}, 
the hidden-charm pentaquarks $P_c$~\cite{LHCb:2015yax,LHCb:2019kea,LHCb:2021chn}, the hidden charm strange pentaquarks $P_{cs}$~\cite{LHCb:2020jpq,LHCb:2022ogu}, the doubly charmed $T_{cc}^+$~\cite{LHCb:2021vvq}, and the excited $\Omega_c$ states~\cite{LHCb:2017uwr} among others. 

These discoveries have posed challenges to conventional quark models~\cite{GellMann:1964nj,Jaffe:1976ig,Esposito:2016noz,Olsen:2017bmm,Lebed:2016hpi} and inspired a multitude of theoretical interpretations, including compact multiquark configurations~\cite{Maiani:2004vq,Ebert:2005nc,Ali:2017jda,Maiani:2014aja}, 
triangle singularities~\cite{Guo:2019twa,Bayar:2016ftu,Liu:2015fea,Aceti:2015zva}, and hadronic molecular models~\cite{Tornqvist:1993ng,Weinstein:1990gu,Karliner:2015ina} have significantly enriched the spectrum of hadronic matter~\cite{Chen:2022asf}.
Among these, the molecular picture—where near-threshold states are generated dynamically via meson-baryon or meson-meson interactions—has been rather successful~\cite{Kaiser:1995eg,Oset:1997it,Oller:2000fj,Jido:2003cb,Hyodo:2011ur,Sun:2011uh,Yang:2011wz,Yamaguchi:2011xb,Garcia-Recio:2003ejq,Uchino:2015uha}.
The molecular picture, in which hadrons are dynamically generated from meson-baryon interactions, has emerged as a compelling framework for understanding many observed states~\cite{Oller:2000ma,Guo:2017jvc,Dong:2021bvy,Dong:2021juy,Chen:2016qju,Dong:2017gaw,Ramos:2020bgs,Wang:2020bjt,Marse-Valera:2022khy,Roca:2024nsi,Hofmann:2005sw}. This approach has been successfully applied to describe the $P_c$, $P_{cs}$, and possible $P_{css}$ systems. Theoretical models involving vector meson exchange~\cite{Garzon:2012np,Liang:2014eba,Liang:2014kra}, heavy quark spin symmetry (HQSS)~\cite{Isgur:1989vq,Neubert:1993mb,Ji:2022vdj,Ji:2022uie,Du:2022jjv}, and coupled-channel dynamics~\cite{Oset:1997it,Hyodo:2011ur,Nieves:2019jhp} have been instrumental in these developments.

Theoretical investigations have shown that vector meson exchange can lead to attractive interactions between a heavy meson and a baryon, possibly forming bound states. This mechanism has been successfully employed within the local hidden gauge approach~\cite{Bando:1984ej,Meissner:1987ge,Nagahiro:2008cv}, where the interaction kernels are derived from effective Lagrangians respecting local gauge invariance. The unitized scattering amplitudes are then obtained through the Bethe-Salpeter equation, allowing the identification of dynamically generated states as poles in the complex energy plane~\cite{Oller:1997ti}.

States with hidden bottom and multiple strange quarks remain relatively unexplored, despite their natural emergence in hadron-hadron molecular models.
Exotic systems composed of a bottom meson and a strange baryon are natural candidates for molecular states. Hadron configurations possessing both heavy and strange quark content, allow the interplay between HQSS and SU(3) flavor symmetry to play a prominent role in binding dynamics~\cite{Chen:2016qju,Liu:2019zoy,Brambilla:2019esw,Ali:2017jda,Karliner:2017qhf}. Despite their theoretical appeal, these systems have been much less studied than their charm counterparts. Exploring these systems contributes not only to a better understanding of QCD dynamics but also to the classification of exotic hadrons beyond the quark model.

Motivated by previous theoretical studies~\cite{Hofmann:2005sw,Yan:2023ttx,Yalikun:2021dpk,Song:2025tha}  on the possibility of deeply bound anticharm multistrange molecular states, we investigate the dynamical generation of bottom–strange baryon molecular states arising from the $S-$wave interaction of bottom mesons ($B$, $B_s$) with ground-state baryons belonging to both the octet and decuplet SU(3) flavor representations. This study includes coupled-channel effects involving pseudoscalar and vector mesons, respecting the heavy quark spin symmetry and flavor symmetry constraints.

Compared to earlier SU(4)-based models~\cite{Wu:2010jy,Wu:2010rv}, our work benefits from much experimental information obtained lately, and shows that one only needs an SU(3) projection, which respects heavy quark spin symmetry. This theoretical foundation allows for robust predictions regarding mass spectra, coupling strengths, and decay channels of potential molecular states.
Our aim is to predict the  properties of potential bottom–multistrange bound or resonance states, thereby deepening our understanding of exotic hadron structure and guiding future experimental searches.

We build on the formalism developed in Refs.~\cite{Debastiani:2017ewu,Feijoo:2022rxf,Yang:2021tvc,Nieves:2019jhp,Liang:2017ejq,Liang:2020dxr}, which employs the local hidden gauge symmetry extended to the heavy flavor sector~\cite{Hofmann:2005sw,Mizutani:2006vq,Tolos:2007vh,Garcia-Recio:2008rjt,Romanets:2012hm} to describe the interactions between heavy mesons and baryons via vector meson exchange, while respecting heavy quark spin symmetry (HQSS) and SU(3) flavor symmetry. Using this approach, we construct the interaction kernels and solve the coupled-channel Bethe-Salpeter equation with a regularization scheme consistent with these symmetries~\cite{Oller:2000fj,Hyodo:2011ur,Debastiani:2017ewu,Yang:2011wz}. This method has been successfully applied to describe known states such as the \(\Omega_c\), \(P_{cs}\), and \(P_{css}\), and enables us to systematically identify dynamically generated molecular states as poles in the scattering amplitudes, extracting their properties, including masses, widths, and coupling strengths. This framework provides a systematic and predictive tool to explore bottom meson–strange baryon interactions across different strangeness sectors.

In this work, we focus on the interactions between bottom mesons ($B^{(*)}$, $B^{(*)}_s$) and ground-state baryons from the octet  and decuplet representations, with an emphasis on systems containing multiple strange quarks. We employ the local hidden gauge approach to derive the interaction potentials and solve the Bethe-Salpeter equation in coupled channels to search for dynamically generated states. By identifying poles in the scattering amplitudes and analyzing their couplings to different channels, we search for possible molecular states with strangeness $S=-1$, $-2$, $-3$, and $-4$ in the open-bottom sector.
This work aims to predict novel exotic baryon resonances and identify the most promising channels for their experimental detection.  
Our study thus contributes to the broader effort of mapping the exotic hadron spectrum by extending molecular state investigations into the bottom and multistrange sectors, an area that remains largely unexplored.

By systematically investigating the bottom–octet and bottom–decuplet sectors, our work expands the molecular state framework to the relatively unexplored multistrange systems, providing robust predictions to guide future experimental efforts at hadron facilities. This study contributes to the understanding of heavy-flavor hadron dynamics and  the comprehensive mapping of exotic hadron structures.

\section{Formalism}
\label{sec:forma}

We study the coupled-channel interactions between bottom mesons, including both $B^{(*)}$ and $B_s^{(*)}$ states, and ground-state baryons from both the octet and decuplet representations. These channels are organized according to their strangeness and isospin quantum numbers, as summarized in Table~\ref{tab:coupled_channels_all}.
\begin{table}[htbp]
\centering
\caption{Coupled channels involving bottom mesons and baryons, classified by baryon type, strangeness ($S$), and isospin ($I$).}
\setlength{\tabcolsep}{36pt}
\begin{tabular}{c|c|c|l}
\hline
\textbf{Baryon Type} & \textbf{Strangeness} & \textbf{Isospin} & \textbf{Channels} \\
\hline
\multirow{5}{*}{Octet} 
  & $S = -1$ & $I = 1/2$ & $B^{(*)}_s N,\ B^{(*)} \Lambda,\ B^{(*)} \Sigma$ \\
  &          & $I = 3/2$ & $B^{(*)} \Sigma$ \\
  & $S = -2$ & $I = 0$   & $B^{(*)}_s \Lambda,\ B^{(*)} \Xi$ \\
  &          & $I = 1$   & $B^{(*)} \Xi$ \\
  & $S = -3$ & $I = 1/2$ & $B^{(*)}_s \Xi$ \\
\hline
\multirow{6}{*}{Decuplet}
  & $S = -1$ & $I = 1/2$ & $B^{(*)} \Sigma^*$ \\
  &          & $I = 3/2$ & $\ B^{(*)}_s \Delta,~B^{(*)} \Sigma^*$ \\
  & $S = -2$ & $I = 0$   & $B^{(*)} \Xi^*$ \\
  &          & $I = 1$   & $B^{(*)}_s \Sigma^*,\ B^{(*)} \Xi^*$ \\
  & $S = -3$ & $I = 1/2$ & $B^{(*)}_s \Xi^*,\ B^{(*)} \Omega$ \\
  & $S = -4$ & $I = 0$   & $B^{(*)}_s \Omega$ \\
\hline
\end{tabular}
\label{tab:coupled_channels_all}
\end{table}

\noindent Together, these four sectors cover a range of strangeness values from $S=-1$ to $S=-4$, allowing us to explore the interactions of bottom mesons with baryons across increasing strangeness.

Our interaction model is constructed on the basis of vector meson exchange, following the local hidden gauge formalism~\cite{Bando:1984ej,bando1988nonlinear,Harada:2003jx,Meissner:1987ge,Nagahiro:2008cv} and extended to the bottom sector in previous works~\cite{Dias:2019klk,Yu:2018yxl,Yu:2019yfr,Liang:2017ejq,Dong:2021bvy,Dong:2021juy,Kong:2021ohg}.

\begin{figure}[H]
  \centering
  \includegraphics[width=4.5cm]{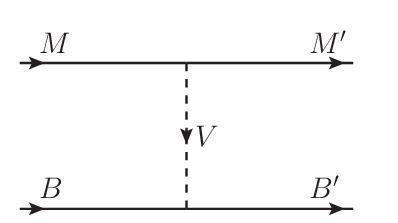}
  \caption{Diagrammatic representation of the meson-baryon interaction $MB \to M^\prime B^\prime$ via vector meson exchange. Here, $M$ ($M^\prime$) and $B$ ($B^\prime$) denote the initial (final) mesons and baryons, respectively, and $V$ represents the exchanged vector meson.}
  \label{fig1}
\end{figure}

As previously mentioned, we employ vector meson exchange based on an extended local hidden gauge approach for the interaction between mesons and baryons, as illustrated in Fig.~\ref{fig1}. 
For the states under consideration, there are two types of vertices involving vector and meson fields, denoted as \( VMM' \) (upper vertex in  Fig.~\ref{fig1}): the \( VPP \) vertex (where \( V \) stands for vector meson and \( P \) for pseudoscalar meson) and the \( VVV \) vertex. These vertices are described by the following Lagrangians~\cite{Nagahiro:2008cv}:
\begin{align}
    \mathcal{L}_{\mathrm{VPP}} &= -i g \left\langle \left[P, \partial_{\mu} P\right] V^{\mu} \right\rangle, \label{eq-1}\\
    \mathcal{L}_{\mathrm{VVV}} &= i g \left\langle \left( V^{\mu} \partial_{\nu} V_{\mu} - \partial_{\nu} V^{\mu} V_{\mu} \right) V^{\nu} \right\rangle. \label{eq-2}
\end{align}

The coupling constant \( g \) is  
$g = \frac{m_V}{2 f_\pi}$, where \( m_V = 800\,\mathrm{MeV} \) is an averaged vector meson mass and \( f_\pi = 93\,\mathrm{MeV} \) is the pion decay constant. Here, \( P \) and \( V \) represent \( q_i \bar{q}_j \)  matrices written in terms of the physical pseudoscalar or vector mesons, respectively. The symbol \(\langle \cdots \rangle\) denotes the trace over these matrices. Although these \( q_i \bar{q}_j \) matrices formally belong to SU(4) or SU(5) flavor groups, the vertices in Eqs.~(\ref{eq-1}) and (\ref{eq-2}) involve only the light  quarks and the $\bar b$ quark is a spectator. This is because the baryons only involve light quarks and hence only light vector,  $\rho,~\omega,~\phi,~\bar K^* $ are allowed to be exchanged. Therefore, invoking full SU(4) or SU(5) symmetry is unnecessary, as discussed in Ref.~\cite{Sakai:2017avl}.

The pseudoscalar ($P$) and vector ($V$) meson matrices employed in our calculations are defined as:
\begin{equation}
 \label{eq:matP_bottom}
    P = \begin{pmatrix}
            \frac{1}{\sqrt{2}}\pi^0 + \frac{1}{\sqrt{3}} \eta + \frac{1}{\sqrt{6}}\eta' & \pi^+ & K^+ & B^+ \\
            \pi^- & -\frac{1}{\sqrt{2}}\pi^0 + \frac{1}{\sqrt{3}} \eta + \frac{1}{\sqrt{6}}\eta' & K^0 & B^0 \\
            K^- & \bar{K}^0 & -\frac{1}{\sqrt{3}} \eta + \sqrt{\frac{2}{3}}\eta' & B_s^0 \\
            B^-  & \bar B^0 & \bar B_s^0 & \eta_b
         \end{pmatrix},
\end{equation}
where the $\eta-\eta'$   mixing of Ref.~\cite{Bramon:1992kr} is used,
\begin{equation}
\label{eq:matV_bottom}
   V = \begin{pmatrix}
            \frac{1}{\sqrt{2}}\rho^0 + \frac{1}{\sqrt{2}} \omega & \rho^+ & K^{* +} & B^{* +} \\
            \rho^- & -\frac{1}{\sqrt{2}}\rho^0 + \frac{1}{\sqrt{2}} \omega  & K^{* 0} & B^{* 0} \\
           K^{* -} & \bar{K}^{* 0}  & \phi & B_s^{* 0} \\
           B^{* -} & \bar B^{* 0} & \bar B_s^{* 0} & \Upsilon
         \end{pmatrix}.
\end{equation}

For the lower vertex of Fig.~\ref{fig1} involving the $VBB'$ coupling we use the Lagrangian 
\begin{equation}
\mathcal{L}_{BBV'} = g\left( \langle \bar{B}\gamma_{\mu}[V'^{\mu},B] \rangle + \langle \bar{B}\gamma_{\mu}B \rangle \langle V'^{\mu} \rangle \right),
\label{lbbv}
\end{equation}
when dealing with baryons of the SU(3) octet, while explicit quark wave functions for the vectors and the baryons are used for the case of baryons of the decuplet~\cite{Song:2025tha,Debastiani:2017ewu}.

The interaction potential $V_{ij}$ corresponding to the diagram in Fig.~\ref{fig1} has the general form:
\begin{equation}
  \label{eq:potential}
  V_{ij} = g^2 (k^0 + k^{'0}) C_{ij},
\end{equation}
where, $ k^0$ is the energy of the initial state meson, and $ k^{'0}$ is  energy of the final state meson, $C_{ij}$ are the dimensionless coupling coefficients. Details on the lower vertex for the vector–baryon coupling can be found in Ref.~\cite{Song:2025tha}, and will not be repeated here.

The combination of the Lagrangian of Eqs.~(\ref{eq-1})~(\ref{eq-2}) with that of  Eq.~(\ref{lbbv}) for baryons of the octet, or explicit quark wave functions for the baryons of the decuplet, allow one to determine the $C_{ij}$ coefficients. They are the same as those found in Ref.~\cite{Song:2025tha} replacing $\bar D$ by $B$ ($\bar c$ quark by 
$\bar b$  quark).
In this section, we present the results for the interactions involving the baryon octet. The coefficients are organized as follows:

\subsection{Pesudoscalar-octet baryon states}
For the sector with strangeness \( S = -1 \) and isospin \( I = 1/2 \), the relevant channels are \( B_s N \), \( B \Lambda \), and \( B \Sigma \). The corresponding coefficients are listed in Table~\ref{coeff_S1}. For completeness, we also quote the coefficient for the \( I = 3/2 \) configuration of the \( B \Sigma \) channel:
\begin{equation}
C(B \Sigma, I = 3/2) = 2.
\end{equation}

\begin{table}[H]
\centering
 \caption{Coefficients $C_{ij}$ for the $S=-1$ sector with isospin $I = 1/2$.}
 \label{coeff_S1}
\setlength{\tabcolsep}{6.5pt}
\begin{tabular}{l|ccc}
\hline
\hline
          ~           & ~$B_s N$~ & ~$B \Lambda$~ & ~$B \Sigma$\\
\hline
$B_s N$~       &        $0$         & $-\sqrt{\frac{3}{2}}$  & $-\sqrt{\frac{3}{2}}$  \\
$B \Lambda$~   &                 &            $1$         &          $0$        \\
$B \Sigma$~    &                 &                     &           $-1$       \\
\hline
\hline
\end{tabular}
\end{table}
In the strangeness \( S = -2 \) sector, we consider the channels \( B_s \Lambda \) and \( B \Xi \) with $I = 0$, with coefficients shown in Table~\ref{coeff_S2}. Additionally, the coupling strength for the \( B \Xi \) channel in the \( I = 1 \) configuration is given by:
\begin{equation}
C(B \Xi, I = 1) = 1.
\end{equation}

\begin{table}[H]
\centering
 \caption{Coefficients $C_{ij}$ for the $S=-2$ sector with isospin $I = 0$.}
 \label{coeff_S2}
\setlength{\tabcolsep}{6.5pt}
\begin{tabular}{l|cc}
\hline
\hline
          ~                 & ~$B_s \Lambda$~ & ~$B \Xi$\\
\hline
$B_s \Lambda$~       &     $1$                 &       $-\sqrt{3} $      \\
$B \Xi$~             &                      &        $-1$      \\
\hline
\hline
\end{tabular}
\end{table}
For the \( S = -3 \) sector, the only relevant channel is \( B_s \Xi \). The corresponding coupling coefficient for the \( B_s \Xi \) channel in the \( I = 1/2 \) configuration is given by:
\begin{equation}
C(B_s \Xi, I = 1/2) = 2.
\end{equation}

\subsection{Pesudoscalar-decuplet baryon states}
The interaction strength among different meson–baryon channels involving the baryon decuplet are encoded in the coupling coefficients \(C_{ij}\), which depend on the strangeness and isospin quantum numbers. Below, we present these coefficients organized by sectors:

In the strangeness \( S = -1 \) sector, the channels \( B_s \Delta \) and \( B \Sigma^* \) are relevant. The coefficient for the \( B \Sigma^* \) channel in the \( I = 1/2 \) sector is
\begin{equation}
C(B \Sigma^*, I = \tfrac{1}{2}) = -1.
\end{equation}
The matrix of coefficients for the \( I = 3/2 \) sector is given in Table~\ref{coeff_decuplet_S1}.
\begin{table}[H]
\centering
 \caption{Coefficients $C_{ij}$ for the $S=-1$ sector with isospin $I=3/2$.}
 \label{coeff_decuplet_S1}
\setlength{\tabcolsep}{6.5pt}
{
\begin{tabular}{l|cc}
\hline
\hline
          ~           & ~$B_s \Delta$~  & ~$B \Sigma^*$\\
\hline
$B_s \Delta$~       &        $0$         & $\sqrt{{3}}$    \\
$B \Sigma^*$~    &                 &                                $2$       \\
\hline
\hline
\end{tabular}}
\end{table}
For the strangeness \( S = -2 \) sector, the channels \( B_s \Sigma^* \) and \( B \Xi^* \) are considered. The coefficient for the \( B \Xi^* \) channel with \( I = 0 \) is
\begin{equation}
C(B \Xi^*, I = 0) = -1.
\end{equation}
The coefficients for the \( I = 1 \) sector are summarized in Table~\ref{coeff_decuplet_S2}.

\begin{table}[H]
\centering
 \caption{Coefficients $C_{ij}$ for the $S=-2$ sector with isospin $I=1$.}
 \label{coeff_decuplet_S2}
\setlength{\tabcolsep}{6.5pt}
{
\begin{tabular}{l|cc}
\hline
\hline
          ~                 & ~$B_s \Sigma^*$~ & ~$B \Xi^*$\\
\hline
$B_s \Lambda$~       &     $1$                 &       $2 $      \\
$B \Xi$~             &                      &        $1$      \\
\hline
\hline
\end{tabular}
}
\end{table}

In the strangeness \( S = -3 \) sector, the channels \( B_s \Xi^* \) and \( B \Omega \) are involved. The coefficients for the \( I = 1/2 \) sector are listed in Table~\ref{coeff_decuplet_S3}.

\begin{table}[H]
\centering
 \caption{Coefficients $C_{ij}$ for the $S=-3$ sector with isospin $I=1/2$.}
 \label{coeff_decuplet_S3}
\setlength{\tabcolsep}{6.5pt}
{
\begin{tabular}{l|cc}
\hline
\hline
          ~                 & ~$B_s \Xi^*$~ & ~$B \Omega$\\
\hline
$B_s \Lambda$~       &     $2$                 &       $\sqrt{3} $      \\
$B \Xi$~             &                      &        $0$      \\
\hline
\hline
\end{tabular}
}
\end{table}

Finally, for the \( S = -4 \) sector, the only channel is \( B_s \Omega \), with the coupling coefficient given in Table~\ref{coeff_S3_1}.
\begin{table}[H]
\centering
 \caption{Coefficient $C_{ij}$ for the $S=-4$ sector with isospin $I=0$.}
 \label{coeff_S3_1}
\setlength{\tabcolsep}{6.5pt}
\begin{tabular}{l|c}
\hline
\hline
          ~           & ~$B_s \Omega$ \\
\hline
$B_s \Omega$ ~  &        $3 $           \\
\hline
\hline
\end{tabular}
\end{table}

\subsection{Vector-baryon states}

For vector baryon states we take the same configurations as before, replacing $B_{(s)}$   by  $B^*_{(s)}$. The interaction is given by 
\begin{equation}
  \label{eq:potentialVVB}
  V_{ij} = g^2 (k^0 + k^{'0}) C_{ij}\vec{\epsilon}~\vec{\epsilon}~',
\end{equation}
where $\vec{\epsilon}~\vec{\epsilon}~'$ are the polarization vectors of the vector mesons. The factor  $\vec{\epsilon}~\vec{\epsilon}~'$ extra with respect to Eq.~(\ref{eq:potential}) does not play any role in the search for poles, couplings, etc. The coefficients $C_{ij}$ are the same as for the pseudoscalar baryon cases. The other novelty in that, having an $S-$wave interaction, while for the case of pseudoscalar-octet baryons the states have $J^P=1/2^-$, and for pseudoscalar-decuplet baryons $3/2^-$, for vector-octet baryons the states are degenerate in $1/2^-,~3/2^-$, and for vector-decuplet baryons they are degenerate in $1/2^-,~3/2^-,~5/2^-$.

\subsection{Amplitudes, poles, couplings, and compositeness}

After constructing the interaction potential \( V_{ij} \), we obtain the scattering matrix by solving the Bethe-Salpeter equation in matrix form:
\begin{equation}
    T = [1 - V G]^{-1} V,
\end{equation}
where \( G \) is a diagonal matrix containing the meson-baryon loop functions for intermediate states. We use a cutoff regularization with \( q_{\text{max}} = 630\,\mathrm{MeV} \) following Ref.~\cite{Oset:1997it}, and vary it within a reasonable range 
taking into account that one found \( 600\,\mathrm{MeV} \) from the \( P_{cs} \) study~\cite{Feijoo:2022rxf} and \( 650\,\mathrm{MeV} \) from the \(\Omega_c\) study~\cite{Debastiani:2017ewu}—to estimate uncertainties.

The loop function for the \( l \)-th channel is
\begin{equation}\label{eq:Gl}
    G_l(\sqrt{s}) = \int_{|\vec{q}|<q_\mathrm{max}} \frac{d^3 q}{(2\pi)^3} \frac{2 M_l \left(w_l(q) + E_l(q)\right)}{2 w_l(q) E_l(q)} \frac{1}{s - \left(w_l(q) + E_l(q)\right)^2 + i \epsilon},
\end{equation}
where \( M_l \) and \( m_l \) are the baryon and meson masses, and
$w_l(q) = \sqrt{m_l^2 + \vec{q}\,^2}, \quad E_l(q) = \sqrt{M_l^2 + \vec{q}\,^2}$.
The total momentum in the meson-baryon rest frame is \( P = (\sqrt{s}, \mathbf{0}) \). In the energy region of interest, \(\operatorname{Re} G\) is negative.

To find poles on the second Riemann sheet, we continue the loop function for open channels as
\begin{equation}\label{eq:GII}
    G_j^{II} = G_j^{I} + i \frac{2 M_j q}{4\pi \sqrt{s}},
\end{equation}
valid for \( \operatorname{Re}(\sqrt{s}) > m_j + M_j \), where the center-of-mass momentum \( q \) is
\begin{equation}\label{eq:qcm}
    q = \frac{\lambda^{1/2}(s, m_j^2, M_j^2)}{2 \sqrt{s}}, \quad \operatorname{Im}(q) > 0.
\end{equation}

The coupling constants \( g_i \) at the pole \( \sqrt{s_P} \) are extracted from the residue of the amplitude,
\begin{equation}
    T_{ij} \approx \frac{g_i g_j}{\sqrt{s} - \sqrt{s_P}}.
\end{equation}
When the pole is near the real axis, we identify
\[
\sqrt{s_P} \approx M + i \frac{\Gamma}{2},
\]
with \( M \) and \( \Gamma \) being the resonance mass and width. One coupling sign can be fixed arbitrarily and the others are defined relative to it. The product \( g_i G_i^{II} \) gives the wave function at the origin in coordinate space~\cite{Gamermann:2009uq}.

We also evaluate the compositeness \( X_i \), which estimates the molecular component  of channel \( i \) in the system for bound states~\cite{Gamermann:2009uq,Hyodo:2013nka}:
\begin{equation}
    X_i = - g_i^2 \left.\frac{\partial G_i}{\partial \sqrt{s}}\right|_{\sqrt{s_P}}.
\end{equation}
This derivative is taken at the pole position. For resonant states \( X_i \) is generally complex, and it is called a “weight” rather than a probability~\cite{Aceti:2014ala}.

\section{Results}
In this section, we present a comprehensive study of the coupled channels constructed from both octet and decuplet baryons combined with bottom mesons. The channels considered span strangeness sectors from \( S = -1 \) to \( S = -4 \). The corresponding threshold masses for all these channels are listed in Table~\ref{thres}.
\begin{table}[H]
\centering
\caption{Threshold masses (in MeV) for the different  channels with strangeness $S = -1$ to $-4$.}
\label{thres}
\setlength{\tabcolsep}{22pt}
\begin{tabular}{l|ccccc}
\hline \hline
  \multirow{4}*{$S=-1$~}
                        & $B_s N$ & $B \Lambda$ & $B \Sigma$ & ~$B_s \Delta$~  & ~$B \Sigma^*$\\
~ &  $6305.2 $ & $6395.09 $ & $6470.03 $ &  $ {6598.93} $  & $6662.24 $  \\
~& $B^*_s N$ & $B^* \Lambda$ & $B^* \Sigma$ & ~$B^*_s \Delta$~  & ~$B^* \Sigma^*$\\
~ &  $6353.67 $ & $6440.43 $ & $6515.24 $ & $6647.4 $ & $6707.61 $   \\
\hline
  \multirow{4}*{$S=-2$~}
                        & $B_s \Lambda$ & $B \Xi$ & ~$B_s \Sigma^*$~ & ~$B \Xi^*$\\
~ &  $6482.61 $ & $6597.85 $  & $6749.76 $ & $6811.37 $   \\
~& $B^*_s \Lambda$ & $B^* \Xi$ & ~$B^*_s \Sigma^*$~ & ~$B^* \Xi^*$ \\
~ &  $6531.08 $ & $6643.06 $ & $6798.23 $ & $6856.58 $     \\
\hline
\hline
  \multirow{4}*{$S=-3$~}
                        &  $B_s \Xi$ & ~$B_s \Xi^*$~ & ~$B \Omega$\\
~ &  $6685.21 $ & $6898.73  $ & $6952.01$  \\
~&  $B^*_s \Xi$ & ~$B^*_s \Xi^*$~ & ~$B^* \Omega$ \\
~ &  $6733.68 $ & $6947.2 $ & $6997.23$    \\
\hline
\hline
  \multirow{4}*{$S=-4$~}
                        & $B_s \Omega$ \\
~ &  $7039.38 $    \\
~&  $B^*_s \Omega$ \\
~ &  $7087.85 $     \\
\hline
\hline
\end{tabular}
\end{table}

The detailed numerical results for the resonance properties—including pole positions, coupling constants ($g_i$), wave function amplitudes at the origin ($g_i G_i$), and compositeness coefficients ($X_i$)—are summarized in Tables~\ref{tab:results1}, \ref{tab:results2}, \ref{tab:results1_1},  \ref{tab:results2_1}, \ref{tab:results_decuplet550} and \ref{tab:results_decuplet650}. These results cover all studied systems, including both octet and decuplet baryon channels, in the strangeness sectors \( S = -1, I = 1/2,~3/2 \) and \( S = -2, I = 0 \). For each system, calculations are performed using two different cutoff values, \( q_{\max} = 550\,\mathrm{MeV} \) and \( q_{\max} = 650\,\mathrm{MeV} \), to estimate the dependence of the results on the regularization scheme.
Notably, no poles were found in the channels with strangeness \( S = -3 \) and \( S = -4 \), indicating the absence of dynamically generated bound or resonance states in these sectors.

\begin{table}[H]
\centering
\caption{Pole positions (in MeV), coupling constants ($g_i$), wave functions at the origin ($g_i G_i$ in $\text{MeV}$), and compositeness  ($X_i$) for the different channels involving octet baryons in the $I=1/2,~S=-1$ and $I=0,~S=-2$ sectors. The results are obtained with a cutoff value of $q_{\text{max}} = 550$ MeV.}
\label{tab:results1}
\setlength{\tabcolsep}{20pt}
\begin{tabular}{l|cccccc}
\hline \hline
Strangeness & Poles & Properties &   \multicolumn{3}{c}{Channels} & $\sum\limits_{i} X_i$ \\
\hline
  \multirow{8}*{$S=-1,~I=1/2$} & ~  & ~  & $B_s N$ & $B \Lambda$ & $B \Sigma$ \\
~ &  & ~  $g_i$ &    $3.50 $ & $2.72 $ & $5.54 $   \\
 & $\textbf{6298}$   & $g_iG_i$    & $-10.27 $ & $-3.78 $ & $-5.53 $  \\
~ &  & ~  ${X_i}$ &   $0.82 $ & $0.06 $ & $0.12 $  & $~\sim1 $ \\
\cline{4-7}~& & ~&   $B^*_s N$ & $B^* \Lambda$ & $B^* \Sigma$ \\
~ &  & ~  $g_i$ &    $3.60 $ & $2.74 $ & $5.64 $   \\
 & $\textbf{6345}$   & $g_iG_i$   & $-10.32 $ & $-3.83 $ & $-5.65 $  \\
~ &  & ~  ${X_i}$ &    $0.81 $ & $0.06 $ & $0.13 $ & $\sim1 $   \\
\hline
  \multirow{8}*{$S=-2,~I=0$} & ~  & ~  & $B_s \Lambda$ & $B \Xi$ \\
~ &  & ~  $g_i$ &    $2.61 $ & $7.78 $  \\
 & $\textbf{6476 }$   & $g_iG_i$   & $-8.87 $ & $-10.09 $   \\
~ &  & ~  ${X_i}$ &   $0.59 $ & $0.41 $ & & $\sim1 $  \\
\cline{4-7}~& & ~&   $B^*_s \Lambda$ & $B^* \Xi$ \\
~ &  & ~  $g_i$ &    $2.73 $ & $7.96 $ & $ $   \\
 & $\textbf{6523}$   & $g_iG_i$   & $-8.94 $ & $-10.33 $   \\
~ &  & ~  ${X_i}$ &    $0.57 $ & $0.43 $ & & $\sim1 $   \\
\hline
\hline
\end{tabular}
\end{table}


\begin{table}[H]
\centering
\caption{Same as Table~\ref{tab:results1}, but with $q_{\text{max}} = 650$ MeV.}
\label{tab:results2}
\setlength{\tabcolsep}{20pt}
\begin{tabular}{l|cccccc}
\hline \hline
Strangeness & Poles & Properties &   \multicolumn{3}{c}{Channels} & $\sum\limits_{i} X_i$ \\
\hline
  \multirow{8}*{$S=-1,~I=1/2$} & ~  & ~  & $B_s N$ & $B \Lambda$ & $B \Sigma$ \\
~ &  & ~  $g_i$ &    $ 6.20$ & $3.60 $ & $8.59 $   \\
 & $\textbf{6254}$   & $g_iG_i$    & $-14.17 $ & $-5.71 $ & $-10.80 $  \\
~ &  & ~  ${X_i}$ &   $0.63 $ & $0.08 $ & $0.28 $  & $\sim1 $ \\
\cline{4-7}~& & ~&   $B^*_s N$ & $B^* \Lambda$ & $B^* \Sigma$ \\
~ &  & ~  $g_i$ &    $6.27 $ & $3.59 $ & $8.65 $   \\
 & $\textbf{6302}$   & $g_iG_i$   & $-14.08 $ & $-5.71 $ & $-10.85 $  \\
~ &  & ~  ${X_i}$ &    $0.62 $ & $0.08 $ & $0.29 $ & $\sim1 $   \\
\hline
  \multirow{8}*{$S=-2,~I=0$} & ~  & ~  & $B_s \Lambda$ & $B \Xi$ \\
~ &  & ~  $g_i$ &    $4.95 $ & $10.91 $  \\
 & $\textbf{6418}$   & $g_iG_i$   & $-11.42 $ & $-15.99 $   \\
~ &  & ~  ${X_i}$ &   $0.37 $ & $0.62 $ & & $\sim1 $  \\
\cline{4-7}~& & ~&   $B^*_s \Lambda$ & $B^* \Xi$ \\
~ &  & ~  $g_i$ &    $5.02 $ & $10.95 $ & $ $   \\
 & $\textbf{6464}$   & $g_iG_i$   & $-11.32 $ & $-15.99 $   \\
~ &  & ~  ${X_i}$ &    $0.37 $ & $0.63 $ & & $\sim1 $   \\
\hline
\hline
\end{tabular}
\end{table}

Tables~\ref{tab:results1} and \ref{tab:results2} summarize the properties of the dynamically generated bound states in the strangeness sectors \(S=-1,~I=1/2\) and \(S=-2,~I=0\), considering both pseudoscalar-baryon ($PB$) and vector-baryon ($VB$) coupled channels, for two different cutoff values, \(q_{\text{max}} = 550~\mathrm{MeV}\) and \(650~\mathrm{MeV}\), respectively. 

\begin{table}[H]
\centering
\caption{Results for decuplet baryon channels in the $I=1/2,~S=-1$ and $I=0,~S=-2$ sectors with $q_{\text{max}} = 550$ MeV.}
\label{tab:results1_1}
\setlength{\tabcolsep}{22pt}
\begin{tabular}{l|ccc}
\hline \hline
Strangeness & Poles (MeV) & Properties &   {Channels}  \\
\hline
  \multirow{8}*{$S=-1,~I=1/2$} & ~  & ~  &  $B \Sigma^*$  \\
~ &  & ~  $g_i$ &    $ 4.17$     \\
 & $\textbf{6646}$   & $g_iG_i$     & $-13.66 $   \\
~ &  & ~  ${X_i}$ &         $\sim1 $   \\
\cline{2-4}~& & ~&    $B^* \Sigma^*$ \\
~ &  & ~  $g_i$ &    $4.19 $   \\
 & $\textbf{6691 }$   & $g_iG_i$     & $-13.61 $   \\
~ &  & ~  ${X_i}$ &         $\sim1 $   \\
\hline  \multirow{8}*{$S=-2,~I=0$} & ~  & ~  &  $B \Xi^*$   \\
~ &  & ~  $g_i$ &    $4.31 $    \\
 & $\textbf{6791}$   & $g_iG_i$     & $-14.14 $   \\
~ &  & ~  ${X_i}$ &         $\sim1 $   \\
\cline{2-4}~& & ~&    $B^* \Xi^*$   \\
~ &  & ~  $g_i$ &    $4.33 $      \\
 & $\textbf{6837}$   & $g_iG_i$     & $-14.09 $   \\
~ &  & ~  ${X_i}$ &         $\sim1 $   \\
\hline
\hline
\end{tabular}
\end{table}

\begin{table}[H]
\centering
\caption{Same as Table~\ref{tab:results1_1}, but with $q_{\text{max}} = 650$ MeV.}
\label{tab:results2_1}
\setlength{\tabcolsep}{22pt}
\begin{tabular}{l|ccc}
\hline \hline
Strangeness & Poles (MeV) & Properties &   {Channels}  \\
\hline
  \multirow{8}*{$S=-1,~I=1/2$} & ~  & ~  &  $B \Sigma^*$  \\
~ &  & ~  $g_i$ &    $5.61 $   \\
 & $\textbf{6625}$   & $g_iG_i$     & $-18.41 $   \\
~ &  & ~  ${X_i}$ &         $\sim1 $   \\
\cline{2-4}~& & ~&    $B^* \Sigma^*$ \\
~ &  & ~  $g_i$ &    $5.64 $     \\
 & $\textbf{6670 }$   & $g_iG_i$     & $-18.34 $   \\
~ &  & ~  ${X_i}$ &         $\sim1 $   \\
\hline  \multirow{8}*{$S=-2,~I=0$} & ~  & ~  &  $B \Xi^*$   \\
~ &  & ~  $g_i$ &    $5.75 $    \\
 & $\textbf{6768}$   & $g_iG_i$     & $-18.86 $   \\
~ &  & ~  ${X_i}$ &         $\sim1 $   \\
\cline{2-4}~& & ~&    $B^* \Xi^*$   \\
~ &  & ~  $g_i$ &    $5.77 $      \\
 & $\textbf{6813}$   & $g_iG_i$     & $-18.79 $   \\
~ &  & ~  ${X_i}$ &         $\sim1 $   \\
\hline
\hline
\end{tabular}
\end{table}

Similarly, Tables~\ref{tab:results1_1} and \ref{tab:results2_1} present the numerical results for the dynamically generated states in decuplet baryon channels with strangeness \(S=-1, I=1/2\) and \(S=-2, I=0\). The results are shown for two different cutoff values, \(q_{\text{max}}=550~\mathrm{MeV}\) and \(650~\mathrm{MeV}\), respectively.

\subsection{Dynamically Generated States in the $S = -1$ Sector}

In the strangeness \( S = -1 \) sector, we study the coupled-channel interactions involving \(B^{(*)}\) and \(B^{(*)}_s\) mesons with both octet and decuplet baryons. The attractive nature of these interactions leads to the formation of bound states, which manifest as poles in the scattering amplitudes.

In the octet sector with \(S = -1, I = 1/2\), bound states are found in the channels \(B_s N\), \(B \Lambda\), and \(B \Sigma\), as well as their vector counterparts \(B_s^* N\), \(B^* \Lambda\), and \(B^* \Sigma\). The pole positions appear near 6298 MeV ($PB$) and 6345 MeV ($VB$) for a cutoff of \(q_{\text{max}} = 550~\mathrm{MeV}\), corresponding to binding energies of approximately 7 MeV and 8 MeV, respectively. Increasing the cutoff to 650 MeV shifts the poles to 6254 MeV ($PB$) and 6302 MeV ($VB$), enhancing the binding energies to about 51 MeV in both channels. 

In the decuplet sector with \(S = -1, I = 1/2\), bound states emerge in the \(B \Sigma^*\) and \(B^* \Sigma^*\) channels. For a cutoff of \(q_{\text{max}} = 550~\mathrm{MeV}\), the pole positions appear around 6646 MeV ($PB$) and 6691 MeV ($VB$), corresponding to binding energies of approximately 16 MeV. Increasing the cutoff to 650 MeV shifts the poles to 6625 MeV and 6670 MeV, increasing the binding energies to about 37 MeV in both cases.

{
It is worth mentioning that in the case of \(S = -1,~I = 3/2\) from the decuplet, bound states are still found despite the apparently repulsive interaction in this sector. This result originates from the strong coupled-channel effects, which allow the generation of bound states in both the pseudoscalar-baryon and vector-baryon systems. The detailed pole positions, couplings, and compositeness for \(q_{\text{max}} = 550~\mathrm{MeV}\) and \(q_{\text{max}} = 650~\mathrm{MeV}\) are shown in Tables~\ref{tab:results_decuplet550} and~\ref{tab:results_decuplet650}, respectively.
}
\begin{table}[H]
\centering
\caption{Pole positions and channel compositions for the decuplet sector with $S = -1,~I = 3/2$, using a cutoff of $q_{\text{max}} = 550$ MeV. }
\label{tab:results_decuplet550}
\setlength{\tabcolsep}{20pt}
\begin{tabular}{l|cccccc}
\hline \hline
Strangeness & Poles & Properties &   \multicolumn{2}{c}{Channels} & & $\sum\limits_{i} X_i$ \\
\hline
  \multirow{8}*{$S=-1,~I=3/2$} & ~  & ~  & $B_s \Delta$~  & ~$B \Sigma^*$ \\
~ &  & ~  $g_i$ &    $2.46  $ & $-2.47  $  \\
 & $\textbf{6595}$   & $g_iG_i$   & $-9.92  $ & $4.62  $   \\
~ &  & ~  ${X_i}$ &   $0.91  $ & $0.08  $ & & $\sim1 $  \\
\cline{4-7}~& & ~&   $B^*_s \Delta$~  & ~$B^* \Sigma^*$ \\
~ &  & ~  $g_i$ &    $2.53  $ & $-2.49  $ & $ $   \\
 & $\textbf{6643}$   & $g_iG_i$   & $-10.03  $ & $4.72  $   \\
~ &  & ~  ${X_i}$ &    $0.90  $ & $0.09  $ & & $\sim1 $   \\
\hline
\hline
\end{tabular}
\end{table}

\begin{table}[H]
\centering
\caption{Same as Table~\ref{tab:results_decuplet550}, with a higher cutoff $q_{\text{max}} = 650$ MeV.} 
\label{tab:results_decuplet650}
\setlength{\tabcolsep}{20pt}
\begin{tabular}{l|cccccc}
\hline \hline
Strangeness & Poles & Properties &   \multicolumn{2}{c}{Channels} & & $\sum\limits_{i} X_i$ \\
\hline
  \multirow{8}*{$S=-1,~I=3/2$} & ~  & ~  & $B_s \Delta$~  & ~$B \Sigma^*$ \\
~ &  & ~  $g_i$ &    $4.17 $ & $-3.37 $  \\
 & $\textbf{6579}$   & $g_iG_i$   & $-15.30 $ & $7.85 $   \\
~ &  & ~  ${X_i}$ &   $0.84 $ & $0.16 $ & & $\sim1 $  \\
\cline{4-7}~& & ~&   $B^*_s \Delta$~  & ~$B^* \Sigma^*$ \\
~ &  & ~  $g_i$ &    $ 4.23 $ & $-3.36  $ & $ $   \\
 & $\textbf{6627}$   & $g_iG_i$   & $-15.27  $ & $7.88  $   \\
~ &  & ~  ${X_i}$ &    $0.83  $ & $0.16  $ & & $\sim1 $   \\
\hline
\hline
\end{tabular}
\end{table}
{
In Table~\ref{tab:results_decuplet550}, we present the results for \(S = -1, I = 3/2\) in the decuplet sector with a cutoff of \(q_{\text{max}} = 550~\mathrm{MeV}\). Two poles are found at 6595 MeV and 6643 MeV, which correspond to the pseudoscalar-baryon ($PB$) and vector-baryon ($VB$) systems, respectively. 
In Table~\ref{tab:results_decuplet650}, the same analysis is performed with a larger cutoff of \(q_{\text{max}} = 650~\mathrm{MeV}\). The pole positions are shifted to lower energies, 6579 MeV and 6627 MeV, indicating an increase in binding energy. The couplings and \(X_i\) values also change accordingly, with stronger channel contributions from both \(B_s \Delta\) and \(B^*_s \Delta\), further confirming the enhanced bound-state nature of these systems under a stronger interaction kernel. The total compositeness $\sum\limits_{i} X_i$ in both cases confirms the molecular nature of the states.
}

This systematic increase in binding energy with the cutoff suggests that the states become more deeply bound as the regularization scale grows. The similar evolution of $ PB $ and $ VB $ poles indicates comparable dynamics in both sectors. 
By construction, the states obtained are of molecular nature and for single channels the compositeness is 1. In the case of coupled channels the compositeness of each channel reflects the probability of this channel in the wave function and the sum of the compositeness for the different channels is also 1.

This trend indicates that the bound states become more deeply bound as the cutoff increases, while the similar binding behavior in $ PB $ and $ VB $ sectors suggests comparable underlying dynamics. Analysis of the coupling constants \(g_i\) and wave functions at the origin \(g_i G_i\)~\cite{Gamermann:2009uq} shows that the dominant component in the  \(S = -1, I = 1/2\) sector is the \(B_s^{(*)} N\) channel, followed by \(B^{(*)} \Sigma\) and \(B^{(*)} \Lambda\). 

\subsection{Dynamically Generated States in the $S = -2$ Sector}

In the $S = -2$ sector, we explore the coupled channel interactions involving \(B^{(*)}\) and \(B^{(*)}_s\) mesons with both octet and decuplet baryons. The attractive interactions among these channels again lead to dynamically generated states.

Within the octet sector characterized by \(S = -2, I = 0\), bound states arise predominantly from the coupled channels \(B_s \Lambda\) and \(B \Xi\), along with their vector counterparts \(B_s^* \Lambda\) and \(B^* \Xi\). The pole positions show a clear dependence on the cutoff parameter \( q_{\max} \): as \( q_{\max} \) increases from 550 MeV to 650 MeV, the $ PB $ pole shifts from approximately 6476 MeV down to 6418 MeV, and the $ VB $ pole moves from around 6523 MeV to 6464 MeV. Correspondingly, the binding energies grow significantly—from about 6 MeV to 64 MeV in the $ PB $ channel, and from 8 MeV to 67 MeV in the $ VB $ channel—indicating a substantial strengthening of the attractive interaction with the increasing cutoff.

Turning to the decuplet sector with the same strangeness and isospin quantum numbers, bound states are generated in the \(B \Xi^*\) and \(B^* \Xi^*\) channels. The poles initially appear near 6791 MeV and 6837 MeV for \(q_{\text{max}}=550~\mathrm{MeV}\), corresponding to binding energies of approximately 20 MeV and 19 MeV for the pseudoscalar-baryon ($PB$) and vector-baryon ($VB$) channels, respectively. As the cutoff increases to \(q_{\text{max}}=650~\mathrm{MeV}\), the poles shift to 6768 MeV and 6813 MeV, enhancing the binding energies to about 43 MeV in both cases. This increase in binding energy with the larger cutoff indicates that the bound states become more deeply bound, while the similar binding energies in the $ PB $ and $ VB $ sectors suggest comparable underlying dynamics.

An examination of the coupling constants \(g_i\) and the wave functions at the origin \(g_i G_i\) reveals that the dominant contributions come from the \(B_s^{(*)} \Lambda\) and \(B^{(*)} \Xi\) channels in the octet sector, while the decuplet states couple predominantly to the \(B^{(*)} \Xi^*\) channels. The compositeness values \(\sum_i X_i\) are close to unity for all these states, as it should be for states generated from an energy independent potential~\cite{Hyodo:2011qc,Gamermann:2009uq}. Note that while there is some energy dependence from ($k^0+k^{'0}$)  in Eq.~(\ref{eq:potential}), the large mass of the $B$ mesons renders this basically energy independent.

\subsection{Absence of Bound States in $S = -3$ and $S = -4$ Sectors}

For strangeness $S = -3$ and $S = -4$, no bound or resonance states are found in our analysis. This is consistent with the fact that all coefficients in the corresponding $C_{ij}$ matrices are positive (see details in section~\ref{sec:forma}), leading to repulsive or non-attractive interactions that prevent pole formation.

Overall, the attractive dynamics between pseudoscalar or vector mesons and strange baryons in the \(S=-1,~I=1/2\) and \(S=-2,~I=0\) sectors results in the appearance of poles in the scattering amplitudes, indicating the formation of bound states. These poles arise from the interactions of \(B^{(*)}\) and \(B_s^{(*)}\) mesons with both octet and decuplet baryons. 
In contrast, for the \(S=-1,~I=3/2\) octet and \(S=-2,~I=1\) sectors, no bound states are found. The bound states remain stable under variations of the cutoff parameter \(q_{\text{max}}\), exhibiting an increase in binding energy as the cutoff value grows. This trend is consistently observed across both the octet and decuplet channels.
It is remarkable that in the case of  \(S=-1,~I=3/2\) from the decuplet we still get bound states in spite of an apparent repulsive interaction. This is due to the important effect of the coupled channels.

\section{Conclusion}

In this work, we have investigated the possibility of dynamically generated molecular states arising from the interaction between bottom mesons and multistrange baryons, employing the local hidden gauge approach in combination with the coupled-channel Bethe–Salpeter equation. 
All the states obtained are of exotic nature since they cannot be reduced to a  3 $q$ state through some $\bar qq$  annihilation.
Our study covers the strangeness sectors from \( S = -1 \) to \( S = -4 \), with particular emphasis on the \( S = -1 \) and \( S = -2 \) sectors, where the attractive interactions between pseudoscalar/vector bottom mesons and strange baryons lead to the formation of bound states.

We predict several molecular states appearing as poles below the corresponding meson-baryon thresholds. In the \( S = -1 \) sector, bound states emerge near 6298 MeV (pseudoscalar-baryon, $PB$) and 6345 MeV (vector-baryon, $VB$) for a cutoff \( q_{\max} = 550\,\mathrm{MeV} \), shifting to about 6254 MeV and 6302 MeV at \( q_{\max} = 650\,\mathrm{MeV} \), indicating increased binding with the larger cutoff. These states are predominantly composed of the \(B_s N\), \(B \Lambda\), and \(B \Sigma\) channels. Similarly, in the decuplet sector, bound states in \(B \Sigma^*\) and \(B^* \Sigma^*\) channels show analogous behavior with poles shifting from around 6646 MeV and 6691 MeV to 6625 MeV and 6670 MeV, respectively, as the cutoff increases.  {Interestingly, in the decuplet sector with \(S = -1, I = 3/2\), we find bound states at \(6595~\mathrm{MeV}\) and \(6643~\mathrm{MeV}\) for \(q_{\text{max}} = 550~\mathrm{MeV}\), which shift to \(6579~\mathrm{MeV}\) and \(6627~\mathrm{MeV}\) for \(q_{\text{max}} = 650~\mathrm{MeV}\), indicating strong binding due to coupled-channel effects despite the repulsive interaction.
}

In the \( S = -2 \) sector, bound states mainly originate from the coupled channels \(B_s \Lambda\), \(B \Xi\), \(B^*_s \Lambda\), and \(B^* \Xi\) in the octet sector, and \(B \Xi^*\), \(B^* \Xi^*\) in the decuplet sector. The $ PB $ and $ VB $ poles initially appear near 6476 MeV and 6523 MeV (octet) and 6791 MeV and 6837 MeV (decuplet) for \(q_{\max} = 550\,\mathrm{MeV}\), shifting to lower energies at \(q_{\max} = 650\,\mathrm{MeV}\), which corresponds to a substantial increase in binding energies (from a few MeV up to ~60 MeV or more).

No bound states are found in the \( S = -1,~I = 3/2 \) {octet} and \( S = -2,~I = 1 \) sectors and in the higher strangeness sectors \( S = -3 \) and \( S = -4 \), regardless of whether the baryon belongs to the octet or decuplet. This can be attributed to the repulsive nature of the interactions in these isospin channels.

Overall, the systematic study across strangeness sectors reveals that the molecular states form  in the \( S=-1, I=1/2 \) and \( S=-2, I=0 \) sectors, with both octet and decuplet baryons. The similar evolution of $ PB $ and $ VB $ states with respect to the cutoff parameter indicates comparable underlying dynamics in both pseudoscalar and vector meson interactions with strange baryons. These findings provide a consistent and robust picture of the hadronic molecular states involving bottom mesons and multistrange baryons.

Our results contribute to  the theoretical understanding of exotic hadrons containing a heavy bottom quark and multiple strange quarks, offering concrete predictions to guide future experimental searches.  The predicted states, especially in the \(S = -1\) and \(S = -2\) sectors, constitute experimentally accessible configurations for exploring the molecular nature of exotic hadrons and  provide critical insights into the nonperturbative regime of QCD involving heavy flavors and multistrangeness.

\section*{ACKNOWLEDGMENTS}
This work is partly supported by the National Natural Science
Foundation of China under Grants  No. 12405089 and No. 12247108 and
the China Postdoctoral Science Foundation under Grant
No. 2022M720360 and No. 2022M720359.  Yi-Yao Li is supported in part by the Guangdong Provincial international exchange program for outstanding young talents of scientific research in 2024. 
This work is also supported by the Spanish
Ministerio de Economia y Competitividad (MINECO) and European FEDER 
funds under Contracts No. FIS2017-84038-
C2-1-P B, PID2020- 112777GB-I00, and by Generalitat Valenciana under 
con- tract PROMETEO/2020/023. This project
has received funding from the European Union Horizon 2020 research and 
innovation programme under the program
H2020- INFRAIA-2018-1, grant agreement No. 824093 of the STRONG-2020 
project. This work is supported by the Spanish Ministerio de Ciencia e 
Innovacion (MICINN) under contracts PID2020-112777GB-I00, 
PID2023-147458NB- ´
C21 and CEX2023-001292-S; by Generalitat Valenciana under contracts 
PROMETEO/2020/023 and CIPROM/2023/59.

\bibliography{refs.bib} 

\begin{thebibliography}{94}
\expandafter\ifx\csname natexlab\endcsname\relax\def\natexlab#1{#1}\fi
\expandafter\ifx\csname bibnamefont\endcsname\relax
  \def\bibnamefont#1{#1}\fi
\expandafter\ifx\csname bibfnamefont\endcsname\relax
  \def\bibfnamefont#1{#1}\fi
\expandafter\ifx\csname citenamefont\endcsname\relax
  \def\citenamefont#1{#1}\fi
\expandafter\ifx\csname url\endcsname\relax
  \def\url#1{\texttt{#1}}\fi
\expandafter\ifx\csname urlprefix\endcsname\relax\def\urlprefix{URL }\fi
\providecommand{\bibinfo}[2]{#2}
\providecommand{\eprint}[2][]{\url{#2}}

\bibitem[{\citenamefont{Choi et~al.}(2003)}]{Belle:2003nnu}
\bibinfo{author}{\bibfnamefont{S.~K.} \bibnamefont{Choi}} \bibnamefont{et~al.} (\bibinfo{collaboration}{Belle}), \bibinfo{journal}{Phys. Rev. Lett.} \textbf{\bibinfo{volume}{91}}, \bibinfo{pages}{262001} (\bibinfo{year}{2003}).

\bibitem[{\citenamefont{Aubert et~al.}(2005)}]{BaBar:2005hhc}
\bibinfo{author}{\bibfnamefont{B.}~\bibnamefont{Aubert}} \bibnamefont{et~al.} (\bibinfo{collaboration}{BaBar}), \bibinfo{journal}{Phys. Rev. Lett.} \textbf{\bibinfo{volume}{95}}, \bibinfo{pages}{142001} (\bibinfo{year}{2005}).

\bibitem[{\citenamefont{Ablikim et~al.}(2013)}]{BESIII:2013ris}
\bibinfo{author}{\bibfnamefont{M.}~\bibnamefont{Ablikim}} \bibnamefont{et~al.} (\bibinfo{collaboration}{BESIII}), \bibinfo{journal}{Phys. Rev. Lett.} \textbf{\bibinfo{volume}{110}}, \bibinfo{pages}{252001} (\bibinfo{year}{2013}).

\bibitem[{\citenamefont{Liu et~al.}(2013)}]{Belle:2013yex}
\bibinfo{author}{\bibfnamefont{Z.~Q.} \bibnamefont{Liu}} \bibnamefont{et~al.} (\bibinfo{collaboration}{Belle}), \bibinfo{journal}{Phys. Rev. Lett.} \textbf{\bibinfo{volume}{110}}, \bibinfo{pages}{252002} (\bibinfo{year}{2013}), \bibinfo{note}{[Erratum: Phys.Rev.Lett. 111, 019901 (2013)]}.

\bibitem[{\citenamefont{Chilikin et~al.}(2013)}]{Belle:2013shl}
\bibinfo{author}{\bibfnamefont{K.}~\bibnamefont{Chilikin}} \bibnamefont{et~al.} (\bibinfo{collaboration}{Belle}), \bibinfo{journal}{Phys. Rev. D} \textbf{\bibinfo{volume}{88}}, \bibinfo{pages}{074026} (\bibinfo{year}{2013}).

\bibitem[{\citenamefont{Aaij et~al.}(2014)}]{LHCb:2014zfx}
\bibinfo{author}{\bibfnamefont{R.}~\bibnamefont{Aaij}} \bibnamefont{et~al.} (\bibinfo{collaboration}{LHCb}), \bibinfo{journal}{Phys. Rev. Lett.} \textbf{\bibinfo{volume}{112}}, \bibinfo{pages}{222002} (\bibinfo{year}{2014}).

\bibitem[{\citenamefont{Aaij and others [LHCb~Collaboration]}(2020)}]{LHCb:2020bwg}
\bibinfo{author}{\bibfnamefont{R.}~\bibnamefont{Aaij}} \bibnamefont{and} \bibinfo{author}{\bibnamefont{others [LHCb~Collaboration]}}, \bibinfo{journal}{Phys. Rev. Lett.} \textbf{\bibinfo{volume}{124}}, \bibinfo{pages}{222002} (\bibinfo{year}{2020}).

\bibitem[{\citenamefont{Aaij et~al.}(2015)}]{LHCb:2015yax}
\bibinfo{author}{\bibfnamefont{R.}~\bibnamefont{Aaij}} \bibnamefont{et~al.} (\bibinfo{collaboration}{LHCb}), \bibinfo{journal}{Phys. Rev. Lett.} \textbf{\bibinfo{volume}{115}}, \bibinfo{pages}{072001} (\bibinfo{year}{2015}).

\bibitem[{\citenamefont{Aaij et~al.}(2019)}]{LHCb:2019kea}
\bibinfo{author}{\bibfnamefont{R.}~\bibnamefont{Aaij}} \bibnamefont{et~al.} (\bibinfo{collaboration}{LHCb}), \bibinfo{journal}{Phys. Rev. Lett.} \textbf{\bibinfo{volume}{122}}, \bibinfo{pages}{222001} (\bibinfo{year}{2019}).

\bibitem[{\citenamefont{Aaij et~al.}(2021{\natexlab{a}})}]{LHCb:2021chn}
\bibinfo{author}{\bibfnamefont{R.}~\bibnamefont{Aaij}} \bibnamefont{et~al.} (\bibinfo{collaboration}{LHCb}), \bibinfo{journal}{Phys. Rev. D} \textbf{\bibinfo{volume}{104}}, \bibinfo{pages}{012005} (\bibinfo{year}{2021}{\natexlab{a}}).

\bibitem[{\citenamefont{Aaij et~al.}(2021{\natexlab{b}})}]{LHCb:2020jpq}
\bibinfo{author}{\bibfnamefont{R.}~\bibnamefont{Aaij}} \bibnamefont{et~al.} (\bibinfo{collaboration}{LHCb}), \bibinfo{journal}{Sci. Bull.} \textbf{\bibinfo{volume}{66}}, \bibinfo{pages}{1278} (\bibinfo{year}{2021}{\natexlab{b}}), \eprint{2012.10380}.

\bibitem[{\citenamefont{Aaij et~al.}(2023)}]{LHCb:2022ogu}
\bibinfo{author}{\bibfnamefont{R.}~\bibnamefont{Aaij}} \bibnamefont{et~al.} (\bibinfo{collaboration}{LHCb}), \bibinfo{journal}{Phys. Rev. Lett.} \textbf{\bibinfo{volume}{131}}, \bibinfo{pages}{031901} (\bibinfo{year}{2023}).

\bibitem[{\citenamefont{Aaij et~al.}(2022)}]{LHCb:2021vvq}
\bibinfo{author}{\bibfnamefont{R.}~\bibnamefont{Aaij}} \bibnamefont{et~al.} (\bibinfo{collaboration}{LHCb}), \bibinfo{journal}{Nature Phys.} \textbf{\bibinfo{volume}{18}}, \bibinfo{pages}{751} (\bibinfo{year}{2022}).

\bibitem[{\citenamefont{Aaij et~al.}(2017)}]{LHCb:2017uwr}
\bibinfo{author}{\bibfnamefont{R.}~\bibnamefont{Aaij}} \bibnamefont{et~al.} (\bibinfo{collaboration}{LHCb}), \bibinfo{journal}{Phys. Rev. Lett.} \textbf{\bibinfo{volume}{118}}, \bibinfo{pages}{182001} (\bibinfo{year}{2017}).

\bibitem[{\citenamefont{Gell-Mann}(1964)}]{GellMann:1964nj}
\bibinfo{author}{\bibfnamefont{M.}~\bibnamefont{Gell-Mann}}, \bibinfo{journal}{Phys. Lett.} \textbf{\bibinfo{volume}{8}}, \bibinfo{pages}{214} (\bibinfo{year}{1964}).

\bibitem[{\citenamefont{Jaffe}(1977)}]{Jaffe:1976ig}
\bibinfo{author}{\bibfnamefont{R.~L.} \bibnamefont{Jaffe}}, \bibinfo{journal}{Phys. Rev. D} \textbf{\bibinfo{volume}{15}}, \bibinfo{pages}{267} (\bibinfo{year}{1977}).

\bibitem[{\citenamefont{Esposito et~al.}(2017)\citenamefont{Esposito, Pilloni, and Polosa}}]{Esposito:2016noz}
\bibinfo{author}{\bibfnamefont{A.}~\bibnamefont{Esposito}}, \bibinfo{author}{\bibfnamefont{A.}~\bibnamefont{Pilloni}}, \bibnamefont{and} \bibinfo{author}{\bibfnamefont{A.~D.} \bibnamefont{Polosa}}, \bibinfo{journal}{Phys. Rept.} \textbf{\bibinfo{volume}{668}}, \bibinfo{pages}{1} (\bibinfo{year}{2017}).

\bibitem[{\citenamefont{Olsen et~al.}(2018)\citenamefont{Olsen, Skwarnicki, and Zieminska}}]{Olsen:2017bmm}
\bibinfo{author}{\bibfnamefont{S.~L.} \bibnamefont{Olsen}}, \bibinfo{author}{\bibfnamefont{T.}~\bibnamefont{Skwarnicki}}, \bibnamefont{and} \bibinfo{author}{\bibfnamefont{D.}~\bibnamefont{Zieminska}}, \bibinfo{journal}{Rev. Mod. Phys.} \textbf{\bibinfo{volume}{90}}, \bibinfo{pages}{015003} (\bibinfo{year}{2018}).

\bibitem[{\citenamefont{Lebed et~al.}(2017)\citenamefont{Lebed, Mitchell, and Swanson}}]{Lebed:2016hpi}
\bibinfo{author}{\bibfnamefont{R.~F.} \bibnamefont{Lebed}}, \bibinfo{author}{\bibfnamefont{R.~E.} \bibnamefont{Mitchell}}, \bibnamefont{and} \bibinfo{author}{\bibfnamefont{E.~S.} \bibnamefont{Swanson}}, \bibinfo{journal}{Prog. Part. Nucl. Phys.} \textbf{\bibinfo{volume}{93}}, \bibinfo{pages}{143} (\bibinfo{year}{2017}).

\bibitem[{\citenamefont{Maiani et~al.}(2005)\citenamefont{Maiani, Piccinini, Polosa, and Riquer}}]{Maiani:2004vq}
\bibinfo{author}{\bibfnamefont{L.}~\bibnamefont{Maiani}}, \bibinfo{author}{\bibfnamefont{F.}~\bibnamefont{Piccinini}}, \bibinfo{author}{\bibfnamefont{A.~D.} \bibnamefont{Polosa}}, \bibnamefont{and} \bibinfo{author}{\bibfnamefont{V.}~\bibnamefont{Riquer}}, \bibinfo{journal}{Phys. Rev. D} \textbf{\bibinfo{volume}{71}}, \bibinfo{pages}{014028} (\bibinfo{year}{2005}).

\bibitem[{\citenamefont{Ebert et~al.}(2006)\citenamefont{Ebert, Faustov, and Galkin}}]{Ebert:2005nc}
\bibinfo{author}{\bibfnamefont{D.}~\bibnamefont{Ebert}}, \bibinfo{author}{\bibfnamefont{R.~N.} \bibnamefont{Faustov}}, \bibnamefont{and} \bibinfo{author}{\bibfnamefont{V.~O.} \bibnamefont{Galkin}}, \bibinfo{journal}{Phys. Lett. B} \textbf{\bibinfo{volume}{634}}, \bibinfo{pages}{214} (\bibinfo{year}{2006}).

\bibitem[{\citenamefont{Ali et~al.}(2017)\citenamefont{Ali, Lange, and Stone}}]{Ali:2017jda}
\bibinfo{author}{\bibfnamefont{A.}~\bibnamefont{Ali}}, \bibinfo{author}{\bibfnamefont{J.~S.} \bibnamefont{Lange}}, \bibnamefont{and} \bibinfo{author}{\bibfnamefont{S.}~\bibnamefont{Stone}}, \bibinfo{journal}{Prog. Part. Nucl. Phys.} \textbf{\bibinfo{volume}{97}}, \bibinfo{pages}{123} (\bibinfo{year}{2017}).

\bibitem[{\citenamefont{Maiani et~al.}(2014)\citenamefont{Maiani, Piccinini, Polosa, and Riquer}}]{Maiani:2014aja}
\bibinfo{author}{\bibfnamefont{L.}~\bibnamefont{Maiani}}, \bibinfo{author}{\bibfnamefont{F.}~\bibnamefont{Piccinini}}, \bibinfo{author}{\bibfnamefont{A.~D.} \bibnamefont{Polosa}}, \bibnamefont{and} \bibinfo{author}{\bibfnamefont{V.}~\bibnamefont{Riquer}}, \bibinfo{journal}{Phys. Rev. D} \textbf{\bibinfo{volume}{89}}, \bibinfo{pages}{114010} (\bibinfo{year}{2014}).

\bibitem[{\citenamefont{Guo et~al.}(2019)\citenamefont{Guo, Jing, Meißner, and Sakai}}]{Guo:2019twa}
\bibinfo{author}{\bibfnamefont{F.-K.} \bibnamefont{Guo}}, \bibinfo{author}{\bibfnamefont{H.-J.} \bibnamefont{Jing}}, \bibinfo{author}{\bibfnamefont{U.-G.} \bibnamefont{Meißner}}, \bibnamefont{and} \bibinfo{author}{\bibfnamefont{S.}~\bibnamefont{Sakai}}, \bibinfo{journal}{Phys. Rev. D} \textbf{\bibinfo{volume}{99}}, \bibinfo{pages}{091501} (\bibinfo{year}{2019}).

\bibitem[{\citenamefont{Bayar et~al.}(2016)\citenamefont{Bayar, Aceti, Guo, and Oset}}]{Bayar:2016ftu}
\bibinfo{author}{\bibfnamefont{M.}~\bibnamefont{Bayar}}, \bibinfo{author}{\bibfnamefont{F.}~\bibnamefont{Aceti}}, \bibinfo{author}{\bibfnamefont{F.-K.} \bibnamefont{Guo}}, \bibnamefont{and} \bibinfo{author}{\bibfnamefont{E.}~\bibnamefont{Oset}}, \bibinfo{journal}{Phys. Rev. D} \textbf{\bibinfo{volume}{94}}, \bibinfo{pages}{074039} (\bibinfo{year}{2016}).

\bibitem[{\citenamefont{Liu et~al.}(2016)\citenamefont{Liu, Oset, and Roca}}]{Liu:2015fea}
\bibinfo{author}{\bibfnamefont{X.}~\bibnamefont{Liu}}, \bibinfo{author}{\bibfnamefont{E.}~\bibnamefont{Oset}}, \bibnamefont{and} \bibinfo{author}{\bibfnamefont{L.}~\bibnamefont{Roca}}, \bibinfo{journal}{Phys. Rev. D} \textbf{\bibinfo{volume}{93}}, \bibinfo{pages}{074015} (\bibinfo{year}{2016}).

\bibitem[{\citenamefont{Aceti et~al.}(2015)\citenamefont{Aceti, Dias, and Oset}}]{Aceti:2015zva}
\bibinfo{author}{\bibfnamefont{F.}~\bibnamefont{Aceti}}, \bibinfo{author}{\bibfnamefont{J.~M.} \bibnamefont{Dias}}, \bibnamefont{and} \bibinfo{author}{\bibfnamefont{E.}~\bibnamefont{Oset}}, \bibinfo{journal}{Eur. Phys. J. A} \textbf{\bibinfo{volume}{51}}, \bibinfo{pages}{48} (\bibinfo{year}{2015}).

\bibitem[{\citenamefont{Tornqvist}(1994)}]{Tornqvist:1993ng}
\bibinfo{author}{\bibfnamefont{N.~A.} \bibnamefont{Tornqvist}}, \bibinfo{journal}{Z. Phys. C} \textbf{\bibinfo{volume}{61}}, \bibinfo{pages}{525} (\bibinfo{year}{1994}).

\bibitem[{\citenamefont{Weinstein and Isgur}(1990)}]{Weinstein:1990gu}
\bibinfo{author}{\bibfnamefont{J.~D.} \bibnamefont{Weinstein}} \bibnamefont{and} \bibinfo{author}{\bibfnamefont{N.}~\bibnamefont{Isgur}}, \bibinfo{journal}{Phys. Rev. D} \textbf{\bibinfo{volume}{41}}, \bibinfo{pages}{2236} (\bibinfo{year}{1990}).

\bibitem[{\citenamefont{Karliner and Rosner}(2015)}]{Karliner:2015ina}
\bibinfo{author}{\bibfnamefont{M.}~\bibnamefont{Karliner}} \bibnamefont{and} \bibinfo{author}{\bibfnamefont{J.~L.} \bibnamefont{Rosner}}, \bibinfo{journal}{Phys. Rev. Lett.} \textbf{\bibinfo{volume}{115}}, \bibinfo{pages}{122001} (\bibinfo{year}{2015}).

\bibitem[{\citenamefont{Chen et~al.}(2023)\citenamefont{Chen, Chen, Liu, Liu, and Zhu}}]{Chen:2022asf}
\bibinfo{author}{\bibfnamefont{H.-X.} \bibnamefont{Chen}}, \bibinfo{author}{\bibfnamefont{W.}~\bibnamefont{Chen}}, \bibinfo{author}{\bibfnamefont{X.}~\bibnamefont{Liu}}, \bibinfo{author}{\bibfnamefont{Y.-R.} \bibnamefont{Liu}}, \bibnamefont{and} \bibinfo{author}{\bibfnamefont{S.-L.} \bibnamefont{Zhu}}, \bibinfo{journal}{Rept. Prog. Phys.} \textbf{\bibinfo{volume}{86}}, \bibinfo{pages}{026201} (\bibinfo{year}{2023}).

\bibitem[{\citenamefont{Kaiser et~al.}(1995)\citenamefont{Kaiser, Siegel, and Weise}}]{Kaiser:1995eg}
\bibinfo{author}{\bibfnamefont{N.}~\bibnamefont{Kaiser}}, \bibinfo{author}{\bibfnamefont{P.~B.} \bibnamefont{Siegel}}, \bibnamefont{and} \bibinfo{author}{\bibfnamefont{W.}~\bibnamefont{Weise}}, \bibinfo{journal}{Nucl. Phys. A} \textbf{\bibinfo{volume}{594}}, \bibinfo{pages}{325} (\bibinfo{year}{1995}).

\bibitem[{\citenamefont{Oset and Ramos}(1998)}]{Oset:1997it}
\bibinfo{author}{\bibfnamefont{E.}~\bibnamefont{Oset}} \bibnamefont{and} \bibinfo{author}{\bibfnamefont{A.}~\bibnamefont{Ramos}}, \bibinfo{journal}{Nucl. Phys. A} \textbf{\bibinfo{volume}{635}}, \bibinfo{pages}{99} (\bibinfo{year}{1998}).

\bibitem[{\citenamefont{Oller and Meissner}(2001)}]{Oller:2000fj}
\bibinfo{author}{\bibfnamefont{J.~A.} \bibnamefont{Oller}} \bibnamefont{and} \bibinfo{author}{\bibfnamefont{U.~G.} \bibnamefont{Meissner}}, \bibinfo{journal}{Phys. Lett. B} \textbf{\bibinfo{volume}{500}}, \bibinfo{pages}{263} (\bibinfo{year}{2001}).

\bibitem[{\citenamefont{Jido et~al.}(2003)\citenamefont{Jido, Oller, Oset, Ramos, and Meissner}}]{Jido:2003cb}
\bibinfo{author}{\bibfnamefont{D.}~\bibnamefont{Jido}}, \bibinfo{author}{\bibfnamefont{J.~A.} \bibnamefont{Oller}}, \bibinfo{author}{\bibfnamefont{E.}~\bibnamefont{Oset}}, \bibinfo{author}{\bibfnamefont{A.}~\bibnamefont{Ramos}}, \bibnamefont{and} \bibinfo{author}{\bibfnamefont{U.~G.} \bibnamefont{Meissner}}, \bibinfo{journal}{Nucl. Phys. A} \textbf{\bibinfo{volume}{725}}, \bibinfo{pages}{181} (\bibinfo{year}{2003}).

\bibitem[{\citenamefont{Hyodo and Jido}(2012)}]{Hyodo:2011ur}
\bibinfo{author}{\bibfnamefont{T.}~\bibnamefont{Hyodo}} \bibnamefont{and} \bibinfo{author}{\bibfnamefont{D.}~\bibnamefont{Jido}}, \bibinfo{journal}{Prog. Part. Nucl. Phys.} \textbf{\bibinfo{volume}{67}}, \bibinfo{pages}{55} (\bibinfo{year}{2012}).

\bibitem[{\citenamefont{Sun et~al.}(2011)\citenamefont{Sun, He, Liu, Luo, and Zhu}}]{Sun:2011uh}
\bibinfo{author}{\bibfnamefont{Z.-F.} \bibnamefont{Sun}}, \bibinfo{author}{\bibfnamefont{J.}~\bibnamefont{He}}, \bibinfo{author}{\bibfnamefont{X.}~\bibnamefont{Liu}}, \bibinfo{author}{\bibfnamefont{Z.-G.} \bibnamefont{Luo}}, \bibnamefont{and} \bibinfo{author}{\bibfnamefont{S.-L.} \bibnamefont{Zhu}}, \bibinfo{journal}{Phys. Rev. D} \textbf{\bibinfo{volume}{84}}, \bibinfo{pages}{054002} (\bibinfo{year}{2011}).

\bibitem[{\citenamefont{Yang et~al.}(2012)\citenamefont{Yang, Sun, He, Liu, and Zhu}}]{Yang:2011wz}
\bibinfo{author}{\bibfnamefont{Z.-C.} \bibnamefont{Yang}}, \bibinfo{author}{\bibfnamefont{Z.-F.} \bibnamefont{Sun}}, \bibinfo{author}{\bibfnamefont{J.}~\bibnamefont{He}}, \bibinfo{author}{\bibfnamefont{X.}~\bibnamefont{Liu}}, \bibnamefont{and} \bibinfo{author}{\bibfnamefont{S.-L.} \bibnamefont{Zhu}}, \bibinfo{journal}{Chin. Phys. C} \textbf{\bibinfo{volume}{36}}, \bibinfo{pages}{6} (\bibinfo{year}{2012}).

\bibitem[{\citenamefont{Yamaguchi et~al.}(2011)\citenamefont{Yamaguchi, Ohkoda, Yasui, and Hosaka}}]{Yamaguchi:2011xb}
\bibinfo{author}{\bibfnamefont{Y.}~\bibnamefont{Yamaguchi}}, \bibinfo{author}{\bibfnamefont{S.}~\bibnamefont{Ohkoda}}, \bibinfo{author}{\bibfnamefont{S.}~\bibnamefont{Yasui}}, \bibnamefont{and} \bibinfo{author}{\bibfnamefont{A.}~\bibnamefont{Hosaka}}, \bibinfo{journal}{Phys. Rev. D} \textbf{\bibinfo{volume}{84}}, \bibinfo{pages}{014032} (\bibinfo{year}{2011}).

\bibitem[{\citenamefont{Garcia-Recio et~al.}(2004)\citenamefont{Garcia-Recio, Lutz, and Nieves}}]{Garcia-Recio:2003ejq}
\bibinfo{author}{\bibfnamefont{C.}~\bibnamefont{Garcia-Recio}}, \bibinfo{author}{\bibfnamefont{M.~F.~M.} \bibnamefont{Lutz}}, \bibnamefont{and} \bibinfo{author}{\bibfnamefont{J.}~\bibnamefont{Nieves}}, \bibinfo{journal}{Phys. Lett. B} \textbf{\bibinfo{volume}{582}}, \bibinfo{pages}{49} (\bibinfo{year}{2004}).

\bibitem[{\citenamefont{Uchino et~al.}(2016)\citenamefont{Uchino, Liang, and Oset}}]{Uchino:2015uha}
\bibinfo{author}{\bibfnamefont{T.}~\bibnamefont{Uchino}}, \bibinfo{author}{\bibfnamefont{W.-H.} \bibnamefont{Liang}}, \bibnamefont{and} \bibinfo{author}{\bibfnamefont{E.}~\bibnamefont{Oset}}, \bibinfo{journal}{Eur. Phys. J. A} \textbf{\bibinfo{volume}{52}}, \bibinfo{pages}{43} (\bibinfo{year}{2016}).

\bibitem[{\citenamefont{Oller et~al.}(2000)\citenamefont{Oller, Oset, and Ramos}}]{Oller:2000ma}
\bibinfo{author}{\bibfnamefont{J.~A.} \bibnamefont{Oller}}, \bibinfo{author}{\bibfnamefont{E.}~\bibnamefont{Oset}}, \bibnamefont{and} \bibinfo{author}{\bibfnamefont{A.}~\bibnamefont{Ramos}}, \bibinfo{journal}{Prog. Part. Nucl. Phys.} \textbf{\bibinfo{volume}{45}}, \bibinfo{pages}{157} (\bibinfo{year}{2000}).

\bibitem[{\citenamefont{Guo et~al.}(2018)\citenamefont{Guo, Hanhart, Mei\ss{}ner, Wang, Zhao, and Zou}}]{Guo:2017jvc}
\bibinfo{author}{\bibfnamefont{F.-K.} \bibnamefont{Guo}}, \bibinfo{author}{\bibfnamefont{C.}~\bibnamefont{Hanhart}}, \bibinfo{author}{\bibfnamefont{U.-G.} \bibnamefont{Mei\ss{}ner}}, \bibinfo{author}{\bibfnamefont{Q.}~\bibnamefont{Wang}}, \bibinfo{author}{\bibfnamefont{Q.}~\bibnamefont{Zhao}}, \bibnamefont{and} \bibinfo{author}{\bibfnamefont{B.-S.} \bibnamefont{Zou}}, \bibinfo{journal}{Rev. Mod. Phys.} \textbf{\bibinfo{volume}{90}}, \bibinfo{pages}{015004} (\bibinfo{year}{2018}), \bibinfo{note}{[Erratum: Rev.Mod.Phys. 94, 029901 (2022)]}.

\bibitem[{\citenamefont{Dong et~al.}(2021{\natexlab{a}})\citenamefont{Dong, Guo, and Zou}}]{Dong:2021bvy}
\bibinfo{author}{\bibfnamefont{X.-K.} \bibnamefont{Dong}}, \bibinfo{author}{\bibfnamefont{F.-K.} \bibnamefont{Guo}}, \bibnamefont{and} \bibinfo{author}{\bibfnamefont{B.-S.} \bibnamefont{Zou}}, \bibinfo{journal}{Commun. Theor. Phys.} \textbf{\bibinfo{volume}{73}}, \bibinfo{pages}{125201} (\bibinfo{year}{2021}{\natexlab{a}}).

\bibitem[{\citenamefont{Dong et~al.}(2021{\natexlab{b}})\citenamefont{Dong, Guo, and Zou}}]{Dong:2021juy}
\bibinfo{author}{\bibfnamefont{X.-K.} \bibnamefont{Dong}}, \bibinfo{author}{\bibfnamefont{F.-K.} \bibnamefont{Guo}}, \bibnamefont{and} \bibinfo{author}{\bibfnamefont{B.-S.} \bibnamefont{Zou}}, \bibinfo{journal}{Progr. Phys.} \textbf{\bibinfo{volume}{41}}, \bibinfo{pages}{65} (\bibinfo{year}{2021}{\natexlab{b}}).

\bibitem[{\citenamefont{Chen et~al.}(2016)\citenamefont{Chen, Chen, Liu, and Zhu}}]{Chen:2016qju}
\bibinfo{author}{\bibfnamefont{H.-X.} \bibnamefont{Chen}}, \bibinfo{author}{\bibfnamefont{W.}~\bibnamefont{Chen}}, \bibinfo{author}{\bibfnamefont{X.}~\bibnamefont{Liu}}, \bibnamefont{and} \bibinfo{author}{\bibfnamefont{S.-L.} \bibnamefont{Zhu}}, \bibinfo{journal}{Phys. Rept.} \textbf{\bibinfo{volume}{639}}, \bibinfo{pages}{1} (\bibinfo{year}{2016}).

\bibitem[{\citenamefont{Dong et~al.}(2017)\citenamefont{Dong, Faessler, and Lyubovitskij}}]{Dong:2017gaw}
\bibinfo{author}{\bibfnamefont{Y.}~\bibnamefont{Dong}}, \bibinfo{author}{\bibfnamefont{A.}~\bibnamefont{Faessler}}, \bibnamefont{and} \bibinfo{author}{\bibfnamefont{V.~E.} \bibnamefont{Lyubovitskij}}, \bibinfo{journal}{Prog. Part. Nucl. Phys.} \textbf{\bibinfo{volume}{94}}, \bibinfo{pages}{282} (\bibinfo{year}{2017}).

\bibitem[{\citenamefont{Ramos et~al.}(2020)\citenamefont{Ramos, Feijoo, Llorens, and Monta\~na}}]{Ramos:2020bgs}
\bibinfo{author}{\bibfnamefont{A.}~\bibnamefont{Ramos}}, \bibinfo{author}{\bibfnamefont{A.}~\bibnamefont{Feijoo}}, \bibinfo{author}{\bibfnamefont{Q.}~\bibnamefont{Llorens}}, \bibnamefont{and} \bibinfo{author}{\bibfnamefont{G.}~\bibnamefont{Monta\~na}}, \bibinfo{journal}{Few Body Syst.} \textbf{\bibinfo{volume}{61}}, \bibinfo{pages}{34} (\bibinfo{year}{2020}).

\bibitem[{\citenamefont{Wang et~al.}(2021)\citenamefont{Wang, Chen, and Liu}}]{Wang:2020bjt}
\bibinfo{author}{\bibfnamefont{F.-L.} \bibnamefont{Wang}}, \bibinfo{author}{\bibfnamefont{R.}~\bibnamefont{Chen}}, \bibnamefont{and} \bibinfo{author}{\bibfnamefont{X.}~\bibnamefont{Liu}}, \bibinfo{journal}{Phys. Rev. D} \textbf{\bibinfo{volume}{103}}, \bibinfo{pages}{034014} (\bibinfo{year}{2021}).

\bibitem[{\citenamefont{Mars\'e-Valera et~al.}(2023)\citenamefont{Mars\'e-Valera, Magas, and Ramos}}]{Marse-Valera:2022khy}
\bibinfo{author}{\bibfnamefont{J.~A.} \bibnamefont{Mars\'e-Valera}}, \bibinfo{author}{\bibfnamefont{V.~K.} \bibnamefont{Magas}}, \bibnamefont{and} \bibinfo{author}{\bibfnamefont{A.}~\bibnamefont{Ramos}}, \bibinfo{journal}{Phys. Rev. Lett.} \textbf{\bibinfo{volume}{130}}, \bibinfo{pages}{091903} (\bibinfo{year}{2023}).

\bibitem[{\citenamefont{Roca et~al.}(2024)\citenamefont{Roca, Song, and Oset}}]{Roca:2024nsi}
\bibinfo{author}{\bibfnamefont{L.}~\bibnamefont{Roca}}, \bibinfo{author}{\bibfnamefont{J.}~\bibnamefont{Song}}, \bibnamefont{and} \bibinfo{author}{\bibfnamefont{E.}~\bibnamefont{Oset}}, \bibinfo{journal}{Phys. Rev. D} \textbf{\bibinfo{volume}{109}}, \bibinfo{pages}{094005} (\bibinfo{year}{2024}).

\bibitem[{\citenamefont{Hofmann and Lutz}(2005)}]{Hofmann:2005sw}
\bibinfo{author}{\bibfnamefont{J.}~\bibnamefont{Hofmann}} \bibnamefont{and} \bibinfo{author}{\bibfnamefont{M.~F.~M.} \bibnamefont{Lutz}}, \bibinfo{journal}{Nucl. Phys. A} \textbf{\bibinfo{volume}{763}}, \bibinfo{pages}{90} (\bibinfo{year}{2005}).

\bibitem[{\citenamefont{Garzon and Oset}(2013)}]{Garzon:2012np}
\bibinfo{author}{\bibfnamefont{E.~J.} \bibnamefont{Garzon}} \bibnamefont{and} \bibinfo{author}{\bibfnamefont{E.}~\bibnamefont{Oset}}, \bibinfo{journal}{Phys. Rev. D} \textbf{\bibinfo{volume}{87}}, \bibinfo{pages}{014036} (\bibinfo{year}{2013}).

\bibitem[{\citenamefont{Liang et~al.}(2014{\natexlab{a}})\citenamefont{Liang, Xiao, and Oset}}]{Liang:2014eba}
\bibinfo{author}{\bibfnamefont{W.~H.} \bibnamefont{Liang}}, \bibinfo{author}{\bibfnamefont{C.~W.} \bibnamefont{Xiao}}, \bibnamefont{and} \bibinfo{author}{\bibfnamefont{E.}~\bibnamefont{Oset}}, \bibinfo{journal}{Phys. Rev. D} \textbf{\bibinfo{volume}{89}}, \bibinfo{pages}{054023} (\bibinfo{year}{2014}{\natexlab{a}}).

\bibitem[{\citenamefont{Liang et~al.}(2014{\natexlab{b}})\citenamefont{Liang, Xie, and Oset}}]{Liang:2014kra}
\bibinfo{author}{\bibfnamefont{W.~H.} \bibnamefont{Liang}}, \bibinfo{author}{\bibfnamefont{J.~J.} \bibnamefont{Xie}}, \bibnamefont{and} \bibinfo{author}{\bibfnamefont{E.}~\bibnamefont{Oset}}, \bibinfo{journal}{Eur. Phys. J. C} \textbf{\bibinfo{volume}{74}}, \bibinfo{pages}{3023} (\bibinfo{year}{2014}{\natexlab{b}}).

\bibitem[{\citenamefont{Isgur and Wise}(1989)}]{Isgur:1989vq}
\bibinfo{author}{\bibfnamefont{N.}~\bibnamefont{Isgur}} \bibnamefont{and} \bibinfo{author}{\bibfnamefont{M.~B.} \bibnamefont{Wise}}, \bibinfo{journal}{Phys. Lett. B} \textbf{\bibinfo{volume}{232}}, \bibinfo{pages}{113} (\bibinfo{year}{1989}).

\bibitem[{\citenamefont{Neubert}(1994)}]{Neubert:1993mb}
\bibinfo{author}{\bibfnamefont{M.}~\bibnamefont{Neubert}}, \bibinfo{journal}{Phys. Rept.} \textbf{\bibinfo{volume}{245}}, \bibinfo{pages}{259} (\bibinfo{year}{1994}).

\bibitem[{\citenamefont{Ji et~al.}(2023)\citenamefont{Ji, Dong, Albaladejo, Du, Guo, Nieves, and Zou}}]{Ji:2022vdj}
\bibinfo{author}{\bibfnamefont{T.}~\bibnamefont{Ji}}, \bibinfo{author}{\bibfnamefont{X.-K.} \bibnamefont{Dong}}, \bibinfo{author}{\bibfnamefont{M.}~\bibnamefont{Albaladejo}}, \bibinfo{author}{\bibfnamefont{M.-L.} \bibnamefont{Du}}, \bibinfo{author}{\bibfnamefont{F.-K.} \bibnamefont{Guo}}, \bibinfo{author}{\bibfnamefont{J.}~\bibnamefont{Nieves}}, \bibnamefont{and} \bibinfo{author}{\bibfnamefont{B.-S.} \bibnamefont{Zou}}, \bibinfo{journal}{Sci. Bull.} \textbf{\bibinfo{volume}{68}}, \bibinfo{pages}{688} (\bibinfo{year}{2023}).

\bibitem[{\citenamefont{Ji et~al.}(2022)\citenamefont{Ji, Dong, Albaladejo, Du, Guo, and Nieves}}]{Ji:2022uie}
\bibinfo{author}{\bibfnamefont{T.}~\bibnamefont{Ji}}, \bibinfo{author}{\bibfnamefont{X.-K.} \bibnamefont{Dong}}, \bibinfo{author}{\bibfnamefont{M.}~\bibnamefont{Albaladejo}}, \bibinfo{author}{\bibfnamefont{M.-L.} \bibnamefont{Du}}, \bibinfo{author}{\bibfnamefont{F.-K.} \bibnamefont{Guo}}, \bibnamefont{and} \bibinfo{author}{\bibfnamefont{J.}~\bibnamefont{Nieves}}, \bibinfo{journal}{Phys. Rev. D} \textbf{\bibinfo{volume}{106}}, \bibinfo{pages}{094002} (\bibinfo{year}{2022}).

\bibitem[{\citenamefont{Du et~al.}(2022)\citenamefont{Du, Albaladejo, Guo, and Nieves}}]{Du:2022jjv}
\bibinfo{author}{\bibfnamefont{M.-L.} \bibnamefont{Du}}, \bibinfo{author}{\bibfnamefont{M.}~\bibnamefont{Albaladejo}}, \bibinfo{author}{\bibfnamefont{F.-K.} \bibnamefont{Guo}}, \bibnamefont{and} \bibinfo{author}{\bibfnamefont{J.}~\bibnamefont{Nieves}}, \bibinfo{journal}{Phys. Rev. D} \textbf{\bibinfo{volume}{105}}, \bibinfo{pages}{074018} (\bibinfo{year}{2022}).

\bibitem[{\citenamefont{Nieves et~al.}(2020)\citenamefont{Nieves, Pavao, and Tolos}}]{Nieves:2019jhp}
\bibinfo{author}{\bibfnamefont{J.}~\bibnamefont{Nieves}}, \bibinfo{author}{\bibfnamefont{R.}~\bibnamefont{Pavao}}, \bibnamefont{and} \bibinfo{author}{\bibfnamefont{L.}~\bibnamefont{Tolos}}, \bibinfo{journal}{Eur. Phys. J. C} \textbf{\bibinfo{volume}{80}}, \bibinfo{pages}{22} (\bibinfo{year}{2020}).

\bibitem[{\citenamefont{Bando et~al.}(1985)\citenamefont{Bando, Kugo, Uehara, Yamawaki, and Yanagida}}]{Bando:1984ej}
\bibinfo{author}{\bibfnamefont{M.}~\bibnamefont{Bando}}, \bibinfo{author}{\bibfnamefont{T.}~\bibnamefont{Kugo}}, \bibinfo{author}{\bibfnamefont{S.}~\bibnamefont{Uehara}}, \bibinfo{author}{\bibfnamefont{K.}~\bibnamefont{Yamawaki}}, \bibnamefont{and} \bibinfo{author}{\bibfnamefont{T.}~\bibnamefont{Yanagida}}, \bibinfo{journal}{Phys. Rev. Lett.} \textbf{\bibinfo{volume}{54}}, \bibinfo{pages}{1215} (\bibinfo{year}{1985}).

\bibitem[{\citenamefont{Meissner}(1988)}]{Meissner:1987ge}
\bibinfo{author}{\bibfnamefont{U.~G.} \bibnamefont{Meissner}}, \bibinfo{journal}{Phys. Rept.} \textbf{\bibinfo{volume}{161}}, \bibinfo{pages}{213} (\bibinfo{year}{1988}).

\bibitem[{\citenamefont{Nagahiro et~al.}(2009)\citenamefont{Nagahiro, Roca, Hosaka, and Oset}}]{Nagahiro:2008cv}
\bibinfo{author}{\bibfnamefont{H.}~\bibnamefont{Nagahiro}}, \bibinfo{author}{\bibfnamefont{L.}~\bibnamefont{Roca}}, \bibinfo{author}{\bibfnamefont{A.}~\bibnamefont{Hosaka}}, \bibnamefont{and} \bibinfo{author}{\bibfnamefont{E.}~\bibnamefont{Oset}}, \bibinfo{journal}{Phys. Rev. D} \textbf{\bibinfo{volume}{79}}, \bibinfo{pages}{014015} (\bibinfo{year}{2009}).

\bibitem[{\citenamefont{Oller and Oset}(1997)}]{Oller:1997ti}
\bibinfo{author}{\bibfnamefont{J.~A.} \bibnamefont{Oller}} \bibnamefont{and} \bibinfo{author}{\bibfnamefont{E.}~\bibnamefont{Oset}}, \bibinfo{journal}{Nucl. Phys. A} \textbf{\bibinfo{volume}{620}}, \bibinfo{pages}{438} (\bibinfo{year}{1997}).

\bibitem[{\citenamefont{Liu et~al.}(2019)\citenamefont{Liu, Chen, Chen, Liu, and Zhu}}]{Liu:2019zoy}
\bibinfo{author}{\bibfnamefont{Y.-R.} \bibnamefont{Liu}}, \bibinfo{author}{\bibfnamefont{H.-X.} \bibnamefont{Chen}}, \bibinfo{author}{\bibfnamefont{W.}~\bibnamefont{Chen}}, \bibinfo{author}{\bibfnamefont{X.}~\bibnamefont{Liu}}, \bibnamefont{and} \bibinfo{author}{\bibfnamefont{S.-L.} \bibnamefont{Zhu}}, \bibinfo{journal}{Prog. Part. Nucl. Phys.} \textbf{\bibinfo{volume}{107}}, \bibinfo{pages}{237} (\bibinfo{year}{2019}).

\bibitem[{\citenamefont{Brambilla et~al.}(2020)\citenamefont{Brambilla, Eidelman, Hanhart, Nefediev, Shen, Thomas, Vairo, and Yuan}}]{Brambilla:2019esw}
\bibinfo{author}{\bibfnamefont{N.}~\bibnamefont{Brambilla}}, \bibinfo{author}{\bibfnamefont{S.}~\bibnamefont{Eidelman}}, \bibinfo{author}{\bibfnamefont{C.}~\bibnamefont{Hanhart}}, \bibinfo{author}{\bibfnamefont{A.}~\bibnamefont{Nefediev}}, \bibinfo{author}{\bibfnamefont{C.-P.} \bibnamefont{Shen}}, \bibinfo{author}{\bibfnamefont{C.~E.} \bibnamefont{Thomas}}, \bibinfo{author}{\bibfnamefont{A.}~\bibnamefont{Vairo}}, \bibnamefont{and} \bibinfo{author}{\bibfnamefont{C.-Z.} \bibnamefont{Yuan}}, \bibinfo{journal}{Phys. Rept.} \textbf{\bibinfo{volume}{873}}, \bibinfo{pages}{1} (\bibinfo{year}{2020}).

\bibitem[{\citenamefont{Karliner et~al.}(2018)\citenamefont{Karliner, Rosner, and Skwarnicki}}]{Karliner:2017qhf}
\bibinfo{author}{\bibfnamefont{M.}~\bibnamefont{Karliner}}, \bibinfo{author}{\bibfnamefont{J.~L.} \bibnamefont{Rosner}}, \bibnamefont{and} \bibinfo{author}{\bibfnamefont{T.}~\bibnamefont{Skwarnicki}}, \bibinfo{journal}{Ann. Rev. Nucl. Part. Sci.} \textbf{\bibinfo{volume}{68}}, \bibinfo{pages}{17} (\bibinfo{year}{2018}).

\bibitem[{\citenamefont{Yan et~al.}(2024)\citenamefont{Yan, Peng, and Pavon~Valderrama}}]{Yan:2023ttx}
\bibinfo{author}{\bibfnamefont{M.-J.} \bibnamefont{Yan}}, \bibinfo{author}{\bibfnamefont{F.-Z.} \bibnamefont{Peng}}, \bibnamefont{and} \bibinfo{author}{\bibfnamefont{M.}~\bibnamefont{Pavon~Valderrama}}, \bibinfo{journal}{Phys. Rev. D} \textbf{\bibinfo{volume}{109}}, \bibinfo{pages}{014023} (\bibinfo{year}{2024}).

\bibitem[{\citenamefont{Yalikun and Zou}(2022)}]{Yalikun:2021dpk}
\bibinfo{author}{\bibfnamefont{N.}~\bibnamefont{Yalikun}} \bibnamefont{and} \bibinfo{author}{\bibfnamefont{B.-S.} \bibnamefont{Zou}}, \bibinfo{journal}{Phys. Rev. D} \textbf{\bibinfo{volume}{105}}, \bibinfo{pages}{094026} (\bibinfo{year}{2022}).

\bibitem[{\citenamefont{Song and Oset}(2025)}]{Song:2025tha}
\bibinfo{author}{\bibfnamefont{J.}~\bibnamefont{Song}} \bibnamefont{and} \bibinfo{author}{\bibfnamefont{E.}~\bibnamefont{Oset}} (\bibinfo{year}{2025}), \eprint{2506.09262}.

\bibitem[{\citenamefont{Wu et~al.}(2010)\citenamefont{Wu, Molina, Oset, and Zou}}]{Wu:2010jy}
\bibinfo{author}{\bibfnamefont{J.-J.} \bibnamefont{Wu}}, \bibinfo{author}{\bibfnamefont{R.}~\bibnamefont{Molina}}, \bibinfo{author}{\bibfnamefont{E.}~\bibnamefont{Oset}}, \bibnamefont{and} \bibinfo{author}{\bibfnamefont{B.~S.} \bibnamefont{Zou}}, \bibinfo{journal}{Phys. Rev. Lett.} \textbf{\bibinfo{volume}{105}}, \bibinfo{pages}{232001} (\bibinfo{year}{2010}).

\bibitem[{\citenamefont{Wu et~al.}(2012)\citenamefont{Wu, Zhao, and Zou}}]{Wu:2010rv}
\bibinfo{author}{\bibfnamefont{J.-J.} \bibnamefont{Wu}}, \bibinfo{author}{\bibfnamefont{L.}~\bibnamefont{Zhao}}, \bibnamefont{and} \bibinfo{author}{\bibfnamefont{B.~S.} \bibnamefont{Zou}}, \bibinfo{journal}{Phys. Lett. B} \textbf{\bibinfo{volume}{709}}, \bibinfo{pages}{70} (\bibinfo{year}{2012}).

\bibitem[{\citenamefont{Debastiani et~al.}(2018)\citenamefont{Debastiani, Dias, Liang, and Oset}}]{Debastiani:2017ewu}
\bibinfo{author}{\bibfnamefont{V.~R.} \bibnamefont{Debastiani}}, \bibinfo{author}{\bibfnamefont{J.~M.} \bibnamefont{Dias}}, \bibinfo{author}{\bibfnamefont{W.~H.} \bibnamefont{Liang}}, \bibnamefont{and} \bibinfo{author}{\bibfnamefont{E.}~\bibnamefont{Oset}}, \bibinfo{journal}{Phys. Rev. D} \textbf{\bibinfo{volume}{97}}, \bibinfo{pages}{094035} (\bibinfo{year}{2018}).

\bibitem[{\citenamefont{Feijoo et~al.}(2023)\citenamefont{Feijoo, Wang, Xiao, Wu, Oset, Nieves, and Zou}}]{Feijoo:2022rxf}
\bibinfo{author}{\bibfnamefont{A.}~\bibnamefont{Feijoo}}, \bibinfo{author}{\bibfnamefont{W.-F.} \bibnamefont{Wang}}, \bibinfo{author}{\bibfnamefont{C.-W.} \bibnamefont{Xiao}}, \bibinfo{author}{\bibfnamefont{J.-J.} \bibnamefont{Wu}}, \bibinfo{author}{\bibfnamefont{E.}~\bibnamefont{Oset}}, \bibinfo{author}{\bibfnamefont{J.}~\bibnamefont{Nieves}}, \bibnamefont{and} \bibinfo{author}{\bibfnamefont{B.-S.} \bibnamefont{Zou}}, \bibinfo{journal}{Phys. Lett. B} \textbf{\bibinfo{volume}{839}}, \bibinfo{pages}{137760} (\bibinfo{year}{2023}).

\bibitem[{\citenamefont{Yang et~al.}(2022)\citenamefont{Yang, Wang, Wu, Oka, and Zhu}}]{Yang:2021tvc}
\bibinfo{author}{\bibfnamefont{Z.}~\bibnamefont{Yang}}, \bibinfo{author}{\bibfnamefont{G.-J.} \bibnamefont{Wang}}, \bibinfo{author}{\bibfnamefont{J.-J.} \bibnamefont{Wu}}, \bibinfo{author}{\bibfnamefont{M.}~\bibnamefont{Oka}}, \bibnamefont{and} \bibinfo{author}{\bibfnamefont{S.-L.} \bibnamefont{Zhu}}, \bibinfo{journal}{Phys. Rev. Lett.} \textbf{\bibinfo{volume}{128}}, \bibinfo{pages}{112001} (\bibinfo{year}{2022}).

\bibitem[{\citenamefont{Liang et~al.}(2018)\citenamefont{Liang, Dias, Debastiani, and Oset}}]{Liang:2017ejq}
\bibinfo{author}{\bibfnamefont{W.-H.} \bibnamefont{Liang}}, \bibinfo{author}{\bibfnamefont{J.~M.} \bibnamefont{Dias}}, \bibinfo{author}{\bibfnamefont{V.~R.} \bibnamefont{Debastiani}}, \bibnamefont{and} \bibinfo{author}{\bibfnamefont{E.}~\bibnamefont{Oset}}, \bibinfo{journal}{Nucl. Phys. B} \textbf{\bibinfo{volume}{930}}, \bibinfo{pages}{524} (\bibinfo{year}{2018}).

\bibitem[{\citenamefont{Liang and Oset}(2020)}]{Liang:2020dxr}
\bibinfo{author}{\bibfnamefont{W.-H.} \bibnamefont{Liang}} \bibnamefont{and} \bibinfo{author}{\bibfnamefont{E.}~\bibnamefont{Oset}}, \bibinfo{journal}{Phys. Rev. D} \textbf{\bibinfo{volume}{101}}, \bibinfo{pages}{054033} (\bibinfo{year}{2020}).

\bibitem[{\citenamefont{Mizutani and Ramos}(2006)}]{Mizutani:2006vq}
\bibinfo{author}{\bibfnamefont{T.}~\bibnamefont{Mizutani}} \bibnamefont{and} \bibinfo{author}{\bibfnamefont{A.}~\bibnamefont{Ramos}}, \bibinfo{journal}{Phys. Rev. C} \textbf{\bibinfo{volume}{74}}, \bibinfo{pages}{065201} (\bibinfo{year}{2006}).

\bibitem[{\citenamefont{Tolos et~al.}(2008)\citenamefont{Tolos, Ramos, and Mizutani}}]{Tolos:2007vh}
\bibinfo{author}{\bibfnamefont{L.}~\bibnamefont{Tolos}}, \bibinfo{author}{\bibfnamefont{A.}~\bibnamefont{Ramos}}, \bibnamefont{and} \bibinfo{author}{\bibfnamefont{T.}~\bibnamefont{Mizutani}}, \bibinfo{journal}{Phys. Rev. C} \textbf{\bibinfo{volume}{77}}, \bibinfo{pages}{015207} (\bibinfo{year}{2008}).

\bibitem[{\citenamefont{Garcia-Recio et~al.}(2011)\citenamefont{Garcia-Recio, Nieves, and Tolos}}]{Garcia-Recio:2008rjt}
\bibinfo{author}{\bibfnamefont{C.}~\bibnamefont{Garcia-Recio}}, \bibinfo{author}{\bibfnamefont{J.}~\bibnamefont{Nieves}}, \bibnamefont{and} \bibinfo{author}{\bibfnamefont{L.}~\bibnamefont{Tolos}}, \bibinfo{journal}{Phys. Rev. C} \textbf{\bibinfo{volume}{83}}, \bibinfo{pages}{065207} (\bibinfo{year}{2011}).

\bibitem[{\citenamefont{Romanets et~al.}(2012)}]{Romanets:2012hm}
\bibinfo{author}{\bibfnamefont{O.}~\bibnamefont{Romanets}} \bibnamefont{et~al.}, \bibinfo{journal}{Physical Review D} \textbf{\bibinfo{volume}{85}}, \bibinfo{pages}{114032} (\bibinfo{year}{2012}).

\bibitem[{\citenamefont{Bando et~al.}(1988)\citenamefont{Bando, Kugo, and Yamawaki}}]{bando1988nonlinear}
\bibinfo{author}{\bibfnamefont{M.}~\bibnamefont{Bando}}, \bibinfo{author}{\bibfnamefont{T.}~\bibnamefont{Kugo}}, \bibnamefont{and} \bibinfo{author}{\bibfnamefont{K.}~\bibnamefont{Yamawaki}}, \bibinfo{journal}{Physics Reports} \textbf{\bibinfo{volume}{164}}, \bibinfo{pages}{217} (\bibinfo{year}{1988}).

\bibitem[{\citenamefont{Harada and Yamawaki}(2003)}]{Harada:2003jx}
\bibinfo{author}{\bibfnamefont{M.}~\bibnamefont{Harada}} \bibnamefont{and} \bibinfo{author}{\bibfnamefont{K.}~\bibnamefont{Yamawaki}}, \bibinfo{journal}{Physics Reports} \textbf{\bibinfo{volume}{381}}, \bibinfo{pages}{1} (\bibinfo{year}{2003}).

\bibitem[{\citenamefont{Dias et~al.}(2020)\citenamefont{Dias, Yu, Liang, Sun, Xie, and Oset}}]{Dias:2019klk}
\bibinfo{author}{\bibfnamefont{J.~M.} \bibnamefont{Dias}}, \bibinfo{author}{\bibfnamefont{Q.-X.} \bibnamefont{Yu}}, \bibinfo{author}{\bibfnamefont{W.-H.} \bibnamefont{Liang}}, \bibinfo{author}{\bibfnamefont{Z.-F.} \bibnamefont{Sun}}, \bibinfo{author}{\bibfnamefont{J.-J.} \bibnamefont{Xie}}, \bibnamefont{and} \bibinfo{author}{\bibfnamefont{E.}~\bibnamefont{Oset}}, \bibinfo{journal}{Chin. Phys. C} \textbf{\bibinfo{volume}{44}}, \bibinfo{pages}{064101} (\bibinfo{year}{2020}).

\bibitem[{\citenamefont{Yu et~al.}(2019{\natexlab{a}})\citenamefont{Yu, Pavao, Debastiani, and Oset}}]{Yu:2018yxl}
\bibinfo{author}{\bibfnamefont{Q.~X.} \bibnamefont{Yu}}, \bibinfo{author}{\bibfnamefont{R.}~\bibnamefont{Pavao}}, \bibinfo{author}{\bibfnamefont{V.~R.} \bibnamefont{Debastiani}}, \bibnamefont{and} \bibinfo{author}{\bibfnamefont{E.}~\bibnamefont{Oset}}, \bibinfo{journal}{Eur. Phys. J. C} \textbf{\bibinfo{volume}{79}}, \bibinfo{pages}{167} (\bibinfo{year}{2019}{\natexlab{a}}).

\bibitem[{\citenamefont{Yu et~al.}(2019{\natexlab{b}})\citenamefont{Yu, Dias, Liang, and Oset}}]{Yu:2019yfr}
\bibinfo{author}{\bibfnamefont{Q.-X.} \bibnamefont{Yu}}, \bibinfo{author}{\bibfnamefont{J.~M.} \bibnamefont{Dias}}, \bibinfo{author}{\bibfnamefont{W.-H.} \bibnamefont{Liang}}, \bibnamefont{and} \bibinfo{author}{\bibfnamefont{E.}~\bibnamefont{Oset}}, \bibinfo{journal}{Eur. Phys. J. C} \textbf{\bibinfo{volume}{79}}, \bibinfo{pages}{1025} (\bibinfo{year}{2019}{\natexlab{b}}).

\bibitem[{\citenamefont{Kong et~al.}(2021)\citenamefont{Kong, Zhu, Song, and He}}]{Kong:2021ohg}
\bibinfo{author}{\bibfnamefont{S.-Y.} \bibnamefont{Kong}}, \bibinfo{author}{\bibfnamefont{J.-T.} \bibnamefont{Zhu}}, \bibinfo{author}{\bibfnamefont{D.}~\bibnamefont{Song}}, \bibnamefont{and} \bibinfo{author}{\bibfnamefont{J.}~\bibnamefont{He}}, \bibinfo{journal}{Phys. Rev. D} \textbf{\bibinfo{volume}{104}}, \bibinfo{pages}{094012} (\bibinfo{year}{2021}).

\bibitem[{\citenamefont{Sakai et~al.}(2017)\citenamefont{Sakai, Roca, and Oset}}]{Sakai:2017avl}
\bibinfo{author}{\bibfnamefont{S.}~\bibnamefont{Sakai}}, \bibinfo{author}{\bibfnamefont{L.}~\bibnamefont{Roca}}, \bibnamefont{and} \bibinfo{author}{\bibfnamefont{E.}~\bibnamefont{Oset}}, \bibinfo{journal}{Phys. Rev. D} \textbf{\bibinfo{volume}{96}}, \bibinfo{pages}{054023} (\bibinfo{year}{2017}).

\bibitem[{\citenamefont{Bramon et~al.}(1992)\citenamefont{Bramon, Grau, and Pancheri}}]{Bramon:1992kr}
\bibinfo{author}{\bibfnamefont{A.}~\bibnamefont{Bramon}}, \bibinfo{author}{\bibfnamefont{A.}~\bibnamefont{Grau}}, \bibnamefont{and} \bibinfo{author}{\bibfnamefont{G.}~\bibnamefont{Pancheri}}, \bibinfo{journal}{Phys. Lett. B} \textbf{\bibinfo{volume}{283}}, \bibinfo{pages}{416} (\bibinfo{year}{1992}).

\bibitem[{\citenamefont{Gamermann et~al.}(2010)\citenamefont{Gamermann, Nieves, Oset, and Ruiz~Arriola}}]{Gamermann:2009uq}
\bibinfo{author}{\bibfnamefont{D.}~\bibnamefont{Gamermann}}, \bibinfo{author}{\bibfnamefont{J.}~\bibnamefont{Nieves}}, \bibinfo{author}{\bibfnamefont{E.}~\bibnamefont{Oset}}, \bibnamefont{and} \bibinfo{author}{\bibfnamefont{E.}~\bibnamefont{Ruiz~Arriola}}, \bibinfo{journal}{Phys. Rev. D} \textbf{\bibinfo{volume}{81}}, \bibinfo{pages}{014029} (\bibinfo{year}{2010}).

\bibitem[{\citenamefont{Hyodo}(2013)}]{Hyodo:2013nka}
\bibinfo{author}{\bibfnamefont{T.}~\bibnamefont{Hyodo}}, \bibinfo{journal}{Int. J. Mod. Phys. A} \textbf{\bibinfo{volume}{28}}, \bibinfo{pages}{1330045} (\bibinfo{year}{2013}).

\bibitem[{\citenamefont{Aceti et~al.}(2014)\citenamefont{Aceti, Dai, Geng, Oset, and Zhang}}]{Aceti:2014ala}
\bibinfo{author}{\bibfnamefont{F.}~\bibnamefont{Aceti}}, \bibinfo{author}{\bibfnamefont{L.~R.} \bibnamefont{Dai}}, \bibinfo{author}{\bibfnamefont{L.~S.} \bibnamefont{Geng}}, \bibinfo{author}{\bibfnamefont{E.}~\bibnamefont{Oset}}, \bibnamefont{and} \bibinfo{author}{\bibfnamefont{Y.}~\bibnamefont{Zhang}}, \bibinfo{journal}{Eur. Phys. J. A} \textbf{\bibinfo{volume}{50}}, \bibinfo{pages}{57} (\bibinfo{year}{2014}).

\bibitem[{\citenamefont{Hyodo et~al.}(2012)\citenamefont{Hyodo, Jido, and Hosaka}}]{Hyodo:2011qc}
\bibinfo{author}{\bibfnamefont{T.}~\bibnamefont{Hyodo}}, \bibinfo{author}{\bibfnamefont{D.}~\bibnamefont{Jido}}, \bibnamefont{and} \bibinfo{author}{\bibfnamefont{A.}~\bibnamefont{Hosaka}}, \bibinfo{journal}{Phys. Rev. C} \textbf{\bibinfo{volume}{85}}, \bibinfo{pages}{015201} (\bibinfo{year}{2012}).

\end{thebibliography}
\newpage

\end{document}